%% file: pqm.tex
\documentclass[a4paper,12pt]{article}
\usepackage[english]{babel}
\usepackage[dvips]{graphicx}
\usepackage{rotating}
\usepackage{verbatim}
\usepackage{makeidx}
\usepackage{color}
\usepackage{amsfonts}
\usepackage{amsmath}
\usepackage{cite}
\let\Otemize =\itemize
\let\Onumerate =\enumerate
\let\Oescription =\description
\def\Nospacing{\itemsep=0pt\topsep=0pt\partopsep=0pt\parskip=0pt\parsep=0pt}
\def\Topspac{\vspace{-0.5\baselineskip}}
\def\Botspac{\vspace{-0.2\baselineskip}}
\newenvironment{Itemize}{\Topspac\Otemize\Nospacing}{\endlist\Botspac}
\newenvironment{Enumerate}{\Topspac\Onumerate\Nospacing}{\endlist\Botspac}

\graphicspath{{./pics/}}
\newcommand{\alphas}{\alpha_{\rm s}}
\newcommand{\omegac}{\omega_{\rm c}}
\newcommand{\lsim}{\,{\buildrel < \over {_\sim}}\,}
\newcommand{\gsim}{\,{\buildrel > \over {_\sim}}\,}
\newcommand{\sqrtsNN}{\sqrt{s_{\scriptscriptstyle{{\rm NN}}}}}
\newcommand{\av}[1]{\left\langle #1 \right\rangle}

\newcommand{\gev}{\mathrm{GeV}}
\newcommand{\tev}{\mathrm{TeV}}
\newcommand{\fm}{\mathrm{fm}}

\newcommand{\PbPb}{\mbox{Pb--Pb}}

\renewcommand{\AA}{\mbox{nucleus--nucleus}}
\newcommand{\RAA}{R_{\rm AA}}

\newcommand{\pt}{p_{\rm t}}
\renewcommand{\d}{{\rm d}}

\newcommand{\arxiv}[1]{\mbox{arXiv:#1}}

\title{Leading-particle suppression
       \mbox{in high energy nucleus--nucleus collisions}}

\author{A.~Dainese$^{a,}$\footnote{andrea.dainese@pd.infn.it}~, 
        C.~Loizides$^{b,}$\footnote{loizides@ikf.uni-frankfurt.de}~~and 
        G.~Pai\'c$^{c,}$\footnote{guypaic@nuclecu.unam.mx}
\vspace{0.3cm}\\
\small $^a$ {\it Universit\`a degli Studi di Padova and INFN, 
via Marzolo 8, 35131 Padova, Italy}\\
\small $^b$ {\it Institut f\"ur Kernphysik, 
August-Euler-Str. 6, D-60486 Frankfurt am Main, Germany}\\
\small $^c$ {\it Instituto de Ciencias Nucleares, UNAM, Mexico City, Mexico}} 

\date{\small{18th June 2004}}

\begin{document}        

\maketitle

\begin{abstract}
\noindent
Parton energy loss effects in heavy-ion collisions are studied
with the Monte Carlo program PQM (Parton Quenching Model) 
constructed using the BDMPS quenching weights and a realistic collision 
geometry. 
The merit of the approach is that it contains only one free parameter that 
is tuned to the high-$\pt$ nuclear modification 
factor measured in central \mbox{Au--Au} collisions at $\sqrtsNN=200~\gev$.
Once tuned, the model is consistently applied to all the high-$\pt$
observables at $200~\gev$: the centrality evolution of the nuclear 
modification factor, the suppression of the away-side jet-like correlations,
and the azimuthal anisotropies for these observables. 
Predictions for the leading-particle suppression at 
\mbox{nucleon--nucleon} centre-of-mass energies of $62.4$ and $5500~\gev$
are presented.
The limits of the eikonal approximation in the BDMPS approach, when applied 
to finite-energy partons, are discussed.
\end{abstract}

\clearpage
\input{src/intro.tex}

\input{src/qw.tex}

\input{src/method.tex}
\input{src/results.tex}

\input{src/disc.tex}

\input{src/concl.tex}

\subsection*{Acknowledgements}
The authors gratefully thank U.A.~Wiedemann for help in the formulation 
of the parton-by-parton calculation of the quenching parameters, 
$\omega_c$ and $R$, A.~Morsch for having provided the Glauber model code,
D.~d'Enterria and C.A.~Salgado for useful comments on the manuscript.
Fruitful and stimulating discussions with F.~Antinori, N.~Armesto, 
K.J.~Eskola, H.~Honkanen, ${\rm K.~\check{S}afa\check{r}}$\'\i k,
J.~Schukraft and R.~Stock are acknowledged.

\end{document}

%% file: src/intro.tex
\section{Introduction}
\label{intro}

High-momentum leading-particle suppression 
in nucleus--nucleus (AA) with respect to proton--proton collisions
is regarded as one of the major discoveries at the Relativistic Heavy Ion 
Collider (RHIC), Brookhaven. 
In Au--Au collisions at centre-of-mass energy $\sqrtsNN=200~\gev$ per 
nucleon--nucleon (NN) pair, the two experiments with high transverse 
momentum, $\pt$, capabilities, PHENIX and STAR, have measured: 
\begin{Itemize}
\item the suppression
      of single particles at high $\pt$ ($\gsim 4~\gev$)
      and central pseudorapidity ($|\eta|\lsim 1$),
      quantified via the nuclear modification factor
      \begin{equation}      
         \label{eq:raa}
         R_{\rm AA}(\pt) \equiv 
         \frac{1}{\av{N_{\rm coll}}_{\rm centrality\,class}} \times
         \frac{{\rm d}^2 N_{\rm AA}/{\rm d}\pt{\rm d}\eta}
              {{\rm d}^2 N_{\rm pp}/{\rm d}\pt{\rm d}\eta} \,,  
      \end{equation}
      which would be equal to 
      unity if the AA collision was a mere superposition of $N_{\rm coll}$ 
      independent NN collisions ($N_{\rm coll}$ scaling);
      instead, at high $\pt$ 
      $\RAA$ is found to decrease from peripheral to central 
      events, down to $\approx 0.2$ 
      in head-on collisions~\cite{phenixRAA,starRAA}; 
      the suppression is the same for charged 
      hadrons and neutral pions for $\pt\gsim 5~\gev$; 
\item the disappearance, in central collisions, of jet-like 
      correlations in the azimuthally-opposite side of a high-$\pt$ leading
      particle~\cite{starIAA}; 
\item the absence of such effects in d--Au collisions at the same 
      energy~\cite{phenixdAu,stardAu}.
\end{Itemize}

These observations can be naturally explained in terms of  
attenuation (quenching) of energetic partons produced in initial 
hard scattering processes, as a consequence of 
the interaction with the dense QCD medium 
expected to be formed in high-energy heavy-ion collisions. 
Several theoretical works exist on the subject~\cite{gyulassywang,bdmps,zakharov,gyulassy2,osborne,zhang,wiedemann,carlosurs}. Most of them implement
the idea of parton energy loss due to medium-induced gluon 
radiation.

In our Monte Carlo program PQM (Parton Quenching Model) we combine a 
recent calculation of parton energy loss~\cite{carlosurs} and a 
realistic description of the collision geometry, which was proven 
to play an important role~\cite{drees}. 
Our approach allows to study and
compare to RHIC data the transverse momentum and 
centrality dependence of single-hadron and 
di-hadron correlation suppressions, as well as the `energy-loss induced' 
azimuthal anisotropy of particle production in non-central collisions.
The model has one single parameter that sets the scale of the energy loss.
Once the parameter is fixed on the basis of the data at $\sqrtsNN=200~\gev$,
we scale it to different energies assuming its proportionality to the 
expected volume-density of gluons, as argued in Ref.~\cite{baier}.
We then apply the same approach to calculate 
the nuclear modification factors at intermediate RHIC energy,
$\sqrtsNN=62.4~\gev$, and at LHC energy, $\sqrtsNN=5.5~\tev$.
Since we do not include so-called initial-state effects, such as nuclear 
modification of the parton distribution functions and parton intrinsic
transverse-momentum broadening, we restrict our study to the high-$\pt$ 
region, above $4$--$5~\gev$ at RHIC energies and above $10~\gev$ at LHC 
energy, where these effects are expected to be 
small (less than 10\% on $\RAA$)~\cite{kariheli,vitev}.

%% file: src/qw.tex
\section{Parton energy loss and collision geometry}
\label{partoneloss}

For the calculation of in-medium parton energy loss,
we use the quenching weights
in the multiple soft scattering approximation, which were
derived in Ref.~\cite{carlosurs} 
in the framework of the `BDMPS' (Baier-Dokshitzer-Mueller-Peign\'e-Schiff) 
formalism~\cite{bdmps}.

In a simplified picture, an energetic parton produced in a hard collision 
undergoes, along its path in the dense medium, multiple scatterings
in a Brownian-like motion with mean free path $\lambda$, which decreases
as the medium density increases. In this multiple scattering process, 
the gluons in the parton wave function pick up
transverse momentum $k_{\rm t}$ with respect to its direction 
and they may eventually decohere and be radiated. 

The scale of the energy loss is set by the characteristic energy 
of the radiated gluons
\begin{equation}
 \omegac = \hat{q}\,L^2/2\,,
 \label{eq:omegac}
\end{equation}
which depends on the in-medium path length $L$ of the parton 
and on the BDMPS transport coefficient of the medium,
$\hat{q}$. The transport coefficient is defined as the average 
medium-induced transverse momentum squared transferred to the parton 
per unit path 
length, $\hat{q} = \av{k_{\rm t}^2}_{\rm medium}\big/\lambda$~\cite{carlosurs}.
For a static medium it is time-independent.  

Differently from the original BDMPS calculation~\cite{bdmps}, 
in Ref.~\cite{carlosurs} 
the transverse momentum $k_{\rm t}$ of a radiated
gluon is kinematically bound to be smaller than its energy $\omega$.
The constraint $k_{\rm t}<\omega$ is imposed via the dimensionless quantity 
\begin{equation}
  R =\frac{2\,\omegac^2}{\hat{q}\,L}=\frac{1}{2}\,\hat{q}\,L^3\,,
 \label{eq:kinematicr}
\end{equation}
which relates the scale of $\omega^2$, given by the square of the 
characteristic energy $\omegac^2$, to that of $k_{\rm t}^2$, 
given by $\hat{q}\,L$, as easily seen from the definition of $\hat{q}$.
The BDMPS case corresponds to $R\to\infty$ and it 
can be recovered by considering an infinitely-extended medium 
($L\to\infty$ for fixed, finite, $\omegac$)~\cite{carlosurs}.

The two parameters $\omegac$ and $R$ determine the energy
distribution of radiated gluons, $\omega\,\d I/\d\omega$. 
While $\omegac$ sets the scale of the distribution,
$R$ controls its shape in the region $0<\omega\ll \omegac$, where the 
kinematic bound $k_{\rm t}<\omega$ is relevant.
In the limit $R\to\infty$ the distribution is of the form~\cite{bdmps}:
\begin{equation} 
 \lim_{R\to\infty} \omega\frac{\d I} {\d \omega} \simeq 
 \frac{2\,\alphas\,C_{\rm R}}{\pi}
 \left\{
 \begin{array}{lll}
 \sqrt{\frac{\omegac}{2\omega}} & {\rm for} & \omega<\omegac \\
 \frac{1}{12}\,\left(\frac{\omegac}{\omega}\right)^2 & {\rm for} & \omega\ge\omegac \\
 \end{array}
 \right.
 \label{eq:wdIdw}
\end{equation}
where $C_{\rm R}$ is the QCD coupling factor (Casimir factor) between the 
considered hard parton and the gluons in the medium; it is  
$C_{\rm F}=4/3$ if the parton is a quark and $C_{\rm A}=3$ 
if the parton is a gluon. 

In the eikonal limit of very large parton initial energy $E$ ($E\gg\omegac$),
the integral of the radiated-gluon energy distribution estimates the 
average energy loss of the parton:
\begin{equation}
\label{eq:avdE}
\av{\Delta E}_{R\to\infty} = \lim_{R\to\infty} \int_0^{\infty} \omega \frac{{\rm d}I}{{\rm d}\omega}\,{\rm d}\omega
\propto \alphas\,C_{\rm R}\,\omegac\propto\alphas\,C_{\rm R}\,\hat{q}\,L^2\,.
\end{equation}
Note that, due to the steep fall-off at large $\omega$ in 
Eq.~(\ref{eq:wdIdw}), 
the integral is dominated by the region $\omega<\omegac$.
The average energy loss $\av{\Delta E}$ is:
proportional to $\alphas\,C_{\rm R}$ and, thus, larger by a factor 
$9/4=2.25$ for gluons than for quarks; 
proportional to the transport coefficient of the medium;
proportional to $L^2$; 
independent of the parton initial energy $E$.  
It is a general feature of all parton energy loss 
calculations~\cite{bdmps,gyulassy2,osborne,zhang,carlosurs}
that the radiated-gluon energy distribution $\omega\,\d I/\d\omega$ 
does not depend on $E$.
Depending on how the kinematic bounds are taken into account, the resulting
$\Delta E$ is $E$-independent (BDMPS)~\cite{bdmps} or depends 
logarithmically on $E$~\cite{gyulassy2,osborne,zhang}.
However, there is always a stronger intrinsic dependence of the radiated 
energy on the initial energy, determined by the fact that the former 
cannot be larger than the latter, $\Delta E\leq E$. 
Within the above simplified derivation which agrees with the main features of 
the BDMPS formalism, this kinematic constraint could be partially included by 
truncating the gluon energy distribution $\omega\,\d I/\d\omega$ at 
the parton energy $E$. 
This would give $\av{\Delta E}\propto \alphas\,C_{\rm R}\,\sqrt{\omegac}\,\sqrt{\min(\omegac,E)}$. For $E<\omegac$, we have 
$\av{\Delta E}\propto\alphas\,C_{\rm R}\,\sqrt{\hat{q}}\,\sqrt{E}\,L$: 
the kinematic constraint
turns the $L$-dependence from quadratic to linear\footnote{Different 
approaches~\cite{gyulassy2,osborne,zhang} 
emphasize the quadratic dependence of energy loss on the 
size of the medium down to rather small parton energies.}
and introduces a 
$\sqrt{E}$-dependence. Note that this procedure implements the  
constraint $\omega\leq E$ for the emission of one gluon, but it does not 
prevent from having $\Delta E=\omega_1+\omega_2+...>E$ 
in a multiple-gluon emission. A full theoretical treatment of the finite 
parton energy case in the BDMPS framework is at present not available. 
As we will discuss in Section~\ref{method}, this introduces significant 
uncertainties in our results.

The probability $P(\Delta E)$ that a hard parton radiates the 
energy $\Delta E$ due to scattering in spatially-extended 
QCD matter is known as the quenching weight~\cite{bdmsJHEP}.
In Ref.~\cite{carlosurs} the weights are calculated
on the basis of the BDMPS formalism for quarks and gluons as a
function of the two parameters $\omegac$ and $R$ and they are given as:
\begin{equation}
 \label{eq:pdeltae}
 P(\Delta E;R,\omegac)=p_0(R)\,\delta(\Delta E) + p(\Delta E;R,\omegac)\,.
\end{equation}
The discrete weight 
$p_0\equiv p_0(R)$ is the probability to have no medium-induced gluon 
radiation 
and the continuous weight $p(\Delta E)\equiv p(\Delta E;R,\omegac)$ 
is the probability to radiate an 
energy $\Delta E$, if at least one gluon is radiated. 
In this work we use the quenching weights calculated in Ref.~\cite{carlosurs}
with a fixed value of the strong coupling $\alphas=1/3$. Note that, since 
$\av{\Delta E}\propto\alphas\,\hat{q}$, the dependence on the value of 
$\alphas$ can be largely absorbed in a rescaling of $\hat{q}$. 

It has been shown~\cite{carlosurs} 
that a simple scaling law exists, which translates 
the radiated-gluon energy distribution for an 
expanding medium with a time-decreasing $\hat{q}(t)$ 
into an equivalent distribution for a static medium, with a
time-averaged $\av{\hat{q}}=constant$, using
\begin{equation}
 \label{eq:qscale}
 \av{\hat{q}}= \frac{2}{L^2} \, \int_{\xi_0}^{L+\xi_0}  
 \left( \xi - \xi_0 \right) \, \hat{q}(\xi) \, \d \xi\,,
\end{equation}
where $\xi_0\sim 10^{-1}~\fm\ll L$ 
is the formation time of the expanding system.

Due to the fact that $\hat{q}$ and $L$ are two more intuitively and 
physically meaningful parameters,
in all the previous applications~\cite{carlosurs,dainese,heli}
the natural $(R,\omegac)$-dependence of the quenching weights 
was `translated' into a $(\hat{q},L)$-dependence, via 
Eqs.~(\ref{eq:omegac}) and~(\ref{eq:kinematicr}). The standard approach 
was to fix a value for the transport coefficient, the same for all
produced partons, and then either use a constant length~\cite{carlosurs}
or calculate a different length for each parton according to 
a description of the collision geometry~\cite{dainese,heli}. 
However, this approach is not optimal, because (a) 
there is no unique and exact definition of the in-medium 
path length when a realistic nuclear density profile is considered, 
as pointed out in Ref.~\cite{dainese},
and (b) the medium density 
is not constant over the whole nucleus--nucleus overlap region but 
rather decreasing from the centre to the periphery.

In order to overcome these limitations, 
we adopt a new approach in which the two 
parameters $\omegac$ and $R$ that determine the quenching weights are
computed on a parton-by-parton basis, taking into account 
both the path length and the density profile of the 
matter traversed by the parton.

Starting from Eq.~(\ref{eq:omegac}) and using Eq.~(\ref{eq:qscale})
with a space-point dependent transport coefficient $\hat{q}(\xi)$ 
and a path-length-averaged $\av{\hat{q}}$, 
we define the effective quantity
\begin{equation}
 \label{eq:omegaceff}
 \omegac \left |_{\rm effective}\right. 
 \equiv \frac{1}{2} \av{\hat{q}}\,L^2 
 = \int_0^{\infty} \xi \, \hat{q}(\xi) \, \d \xi\,,
\end{equation}
which on the r.h.s.~does not explicitly depend on $L$. 
For a step-function `density' 
distribution $\hat{q}(\xi)=\hat{q}_0\,\theta(L -\xi)$, 
Eq.~(\ref{eq:omegaceff}) coincides with Eq.~(\ref{eq:omegac}).
Similarly, we define 
\begin{equation}
 \label{eq:lqhateff}
 \av{\hat{q}}L\left|_{\rm effective}\right. 
 \equiv \int_0^{\infty} \hat{q}(\xi) \, \d \xi
\end{equation}
and
\begin{equation}
 \label{eq:kinematicreff}
 R \left |_{\rm effective}\right. \equiv 
  \frac{2\, \left(\omegac\left |_{\rm effective}\right)^2\right.}
       {\av{\hat{q}}L\left |_{\rm effective}\right.}\,.
\end{equation}
Using the definitions in Eqs.~(\ref{eq:omegaceff})--(\ref{eq:kinematicreff})
we incorporate the collision geometry 
in the calculation of parton energy loss via the `local' 
transport coefficient $\hat{q}(\xi)$.

%% file: src/method.tex
\section{Leading-particle suppression procedure}
\label{method}

Within the perturbative QCD collinear factorization framework,
the expression for the production of high-$\pt$ hadrons 
at central rapidity, $y=0$,
in pp collisions (no energy loss) reads:
\begin{equation}
\label{eq:vacuum}
\left.\frac{\d^2\sigma^h}{\d\pt\d y}\right|_{y=0}=
\sum_{a,b,j=\rm q,\overline q,g}\int\d x_a\,\d x_b\,\d z_j\,
f_a(x_a)\,f_b(x_b)\,
\left.\frac{\d^2\hat{\sigma}^{ab\to jX}}{\d p_{{\rm t},j}\d y_j}\right|_{y_j=0}
\frac{D_{h/j}(z_j)}{z_j^2}\,,
\end{equation}
where $f_{a(b)}$ is the parton distribution function for a parton of 
type $a(b)$ carrying the momentum fraction $x_{a(b)}$, 
$\hat{\sigma}^{ab\to jX}$ are the
partonic hard-scattering cross sections and $D_{h/j}(z_j)$ is the 
fragmentation function, i.e. the probability distribution for the parton $j$ 
to fragment into a hadron $h$ with transverse momentum 
$\pt=z_j\,p_{{\rm t},j}$. To simplify the notation, we have dropped the 
dependence of $\hat{\sigma}^{ab\to jX}$ on $\sqrt{s}$ and of $f_{a(b)}$, 
$\hat{\sigma}^{ab\to jX}$
and $D_{h/j}$ on the square of the scale 
(momentum transfer) $Q^2$ of the hard scattering,
usually $Q^2\sim p_{{\rm t},j}^2$. 
Medium-induced parton energy loss is included by modifying 
Eq.~(\ref{eq:vacuum}) to:
\begin{equation}
\label{eq:medium}
\begin{array}{ll}
\displaystyle{\left.\frac{\d^2\sigma_{\rm quenched}^h}{\d\pt\d y}\right|_{y=0} =} &
\displaystyle{\sum_{a,b,j=\rm q,\overline q,g}}
\displaystyle{\int}\d x_a\,\d x_b\,\d\Delta E_j\,\d z_j\, 
f_a(x_a)\,f_b(x_b)\,
\displaystyle{\left.\frac{\d^2\hat{\sigma}^{ab\to jX}}{\d p^{\rm init}_{{\rm t},j}\d y_j}\right|_{y_j=0}}\times \\
&\displaystyle{\times\,
\delta\left(p^{\rm init}_{{\rm t},j}-(p_{{\rm t},j}+\Delta E_j)\right)\,
P(\Delta E_j;R_j,\omega_{{\rm c},j})\,
\frac{D_{h/j}(z_j)}{z_j^2}\,},
\end{array}
\end{equation}
where $P(\Delta E_j;R_j,\omega_{{\rm c},j})$ 
is the energy-loss probability 
distribution, Eq.~(\ref{eq:pdeltae}), for the parton $j$ 
(we will explain in the following how the input parameters $R$ and $\omegac$ 
for a given parton are calculated). 

In PQM we obtain the leading-particle suppression in \AA~collisions 
by calculating the transverse momentum distributions in Eqs.~(\ref{eq:vacuum}) 
and~(\ref{eq:medium}) in a Monte Carlo approach.
The `event loop' that we iterate is the following:
\begin{Enumerate}
\item generation of a parton, quark or gluon, with $\pt>3~\gev$, 
      using the PYTHIA event generator~\cite{pythia} in pp mode with 
      CTEQ\,4L parton distribution functions~\cite{cteq4}; 
      the $\pt$-dependence of the quarks-to-gluons ratio is taken 
      from PYTHIA;
\item determination of the two input parameters, $\omegac$ and $R$, 
      for the calculation of the quenching weights, i.e. the energy-loss
      probability distribution $P(\Delta E)$;
\item sampling of an energy loss $\Delta E$ according to $P(\Delta E)$ 
      and definition of the new parton
      transverse momentum, $\pt-\Delta E$;
\item (independent) fragmentation of the parton to a hadron using the 
      leading-order Kniehl-Kramer-P\"otter (KKP) fragmentation 
      functions~\cite{kkp}. 
\end{Enumerate}
Steps 2 and 3 are explained in detail in the following paragraphs.
Quenched and unquenched $\pt$ distributions are obtained including or 
excluding the third step of the chain. The nuclear modification 
factor $R_{\rm AA}(\pt)$ is given by their ratio.
Our hadrons $\pt$ distribution without
energy loss at $\sqrt{s}=200~\gev$ agrees in shape with that measured for 
neutral pions in pp collisions by PHENIX~\cite{phenixpp}. 

\subsubsection*{Determination of $\omegac$ and $R$}

We define the collision geometry in the $(x,y)$ plane transverse to 
the beam direction $z$, 
in which the centres of two nuclei A and B colliding with an
impact parameter $b$ have coordinates $(-b/2,0)$ and $(b/2,0)$, respectively.
Using the Glauber model~\cite{glauber} to describe the geometry of 
the collision, we assume (a) the distribution of parton production points
in the transverse plane and (b) the transverse density of the medium both 
to be proportional to the $b$-dependent product 
$T_{\rm A}T_{\rm B}(x,y;b)\equiv T_{\rm A}(x,y)\times T_{\rm B}(x,y)$ 
of the thickness functions of the two nuclei. The nuclear 
thickness function is defined as the $z$-integrated Wood-Saxon nuclear density 
profile: $T_{\rm i}(x,y)\equiv\int \d z\,\rho_i^{\rm WS}(x,y,z)$. 
The parameters 
of the Wood-Saxon profile for different nuclei are tabulated from 
data~\cite{atomdata}. Note that $T_{\rm A}T_{\rm B}(x,y;b)$
estimates the transverse density of binary NN collisions, 
$\rho_{\rm coll}(x,y;b)$, modulo the inelastic NN cross section.

We consider only partons produced at central rapidity and assume 
that they propagate in the transverse plane ($E\approx p\approx \pt$). 
For a parton with production point $(x_0,y_0)$ and azimuthal propagation 
direction $(\cos\phi_0,\sin\phi_0)$ ($\phi_0$ is sampled uniformly), 
we define the `local' 
transport coefficient along the path of the parton inside the overlap region 
of the nuclei as:
\begin{equation}
\hat{q}(\xi;b)=k\times 
T_{\rm A}T_{\rm B}(x_0+\xi \cos\phi_0,y_0+\xi \sin\phi_0;b)\,,
\label{eq:qofxi}
\end{equation}
where $k$ is a free parameter (in $\fm$) 
that sets the scale of the transport coefficient (in $\gev^2/\fm$).
We compute the two integrals $I_0$ and $I_1$ (Eqs.~(\ref{eq:lqhateff}) and 
(\ref{eq:omegaceff}))
\begin{equation}
I_n \equiv \int_0^\infty \xi^n\,\hat{q}(\xi;b)\,\d\xi\hspace{1cm}n=0,\,1\,,
\label{eq:In}
\end{equation}
which determine the energy-loss probability distribution $P(\Delta E)$ 
using\footnote{For simplicity, hereafter we drop the subscript 
``effective'' for $\omegac$ and $R$.} 
(see Section~\ref{partoneloss}):
\begin{equation}
\omegac=I_1\hspace{0.5cm}{\rm and}\hspace{0.5cm}R=2\,I_1^2/I_0\,.
\label{eq:wcRfromI0I1}
\end{equation}

Our approach allows a natural extension from central to peripheral 
nucleus--nucleus 
collisions: the idea is to fix the only free parameter, $k$, in order to 
describe the measured nuclear modification factor in central collisions and 
then use the impact parameter ($b$) dependence of the product 
$T_{\rm A}T_{\rm B}(x,y;b)$. We translate, by means of the 
Glauber model, the experimental definition of the centrality classes 
in terms of fractions of the geometrical cross section to a range 
in $b$ and, within such range, we sample, for every loop of the chain 
reported at the beginning of this section, a value of $b$
according to the $b$-dependence of the average number of binary 
collisions, $\d \av{N^{\rm AB}_{\rm coll}}/\d b$.  

\begin{figure}[t!]
  \begin{center}
    \includegraphics[width=\textwidth]{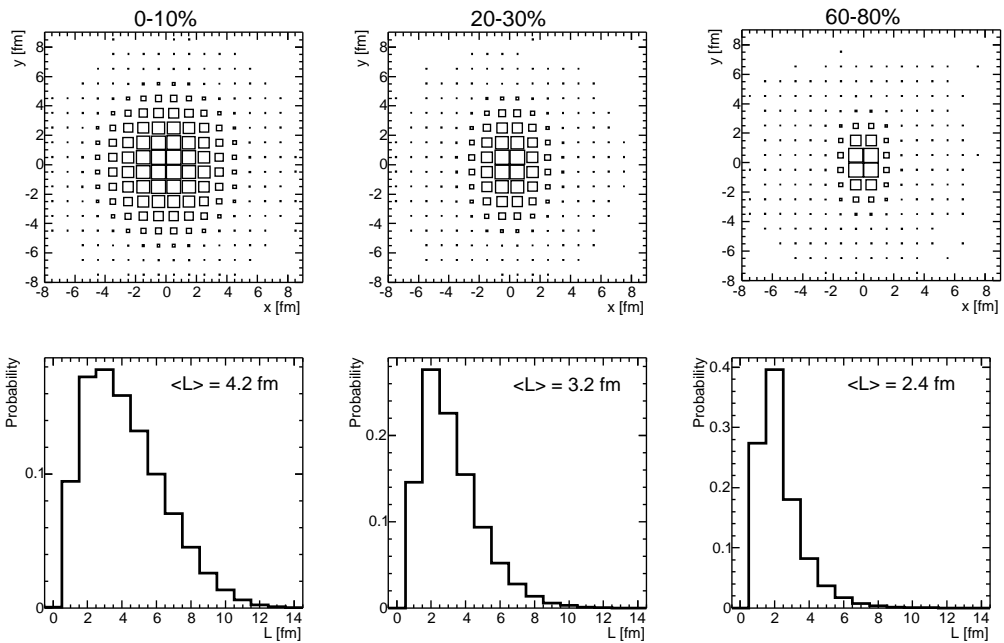}
    \caption{Distributions of parton production points in the transverse plane 
             (upper row) and in-medium path length (lower row) 
             in central, semi-central and peripheral Au--Au collisions. 
             The quantity $\av{L}$ is the average of the path length 
             distribution.} 
    \label{fig:geo}
  \end{center}
\end{figure}

In order to give a synthetic and direct illustration of our results, 
we compute, for a given centrality class, the distributions of the two 
more customary variables $L$ and $\hat{q}$. To this purpose, for every 
parton we combine $\omegac$ and $R$, using Eqs.~(\ref{eq:omegac}) and
(\ref{eq:kinematicr}), to obtain
an effective path length and an effective transport coefficient:
\begin{equation}
\label{eq:L}
L=R/\omegac=2\,I_1/I_0
\hspace{0.5cm}{\rm and}\hspace{0.5cm}
\hat{q}=2\,\omegac^2/(L\,R)=I_0^2/(2\,I_1)\,.
\end{equation}
We point out that the resulting definition of 
$L$ is, as necessary, independent of $k$. Furthermore,
it is the same one of us (A.D.) used in Ref.~\cite{dainese}. Note that 
$\hat{q}$ is proportional to $k$. 
In Fig.~\ref{fig:geo} we report for illustration 
the distributions of parton production points in the transverse plane 
and of in-medium path lengths, in central (0--10\%), semi-central (20--30\%)
and peripheral (60--80\%) Au--Au collisions.
We will show the $\hat{q}$ distributions for different centralities 
in the next section (in Fig.~\ref{fig:QAllCentralities}), after
extracting the scale $k$ from the data.

\subsubsection*{Energy-loss sampling}

In the third step of the chain, 
we use the numerical routine provided in Ref.~\cite{carlosurs} for fixed 
$\alphas=1/3$ to 
obtain the energy-loss probability distribution for given
$\omegac$, $R$ and parton species (quark or gluon). 
According to this
distribution, we sample an energy loss $\Delta E$ to be subtracted from the
parton transverse momentum.
The quenching weights are calculated in the eikonal approximation,
where the energy of the parton is infinite ($E=\pt=\infty$). Therefore, when 
the realistic case of finite-energy partons is considered, a significant 
part of the energy-loss probability distribution $P(\Delta E)$ lies 
above the parton energy $E$, in particular for large values of 
$\omegac$ and $R$, or equivalently, of $\hat{q}$ and $L$. The energy loss,
under the constraints introduced by the finite parton energies, is 
sampled following two approaches:
\begin{Itemize}
\item {\it Reweighted:} truncate $P(\Delta E)$ at $\Delta E=E$ 
and renormalize it to unity by dividing out the factor 
$\int_0^E\d\epsilon\,P(\epsilon)$.
The Monte Carlo implementation of this approach is: sample $\Delta E$
from the original $P(\Delta E)$; if $\Delta E>E$, sample another $\Delta E$;
iterate until a $\Delta E\leq E$ is sampled.
\item {\it Non-reweighted:} truncate $P(\Delta E)$ at $\Delta E=E$ 
and add the $\delta$-function 
$\delta(\Delta E-E)\int_E^\infty\d\epsilon\,P(\epsilon)$
to it. The integral of $P$ is, in this way, maintained equal to one.
The corresponding Monte Carlo implementation reads: sample an energy loss 
$\Delta E$ and set the new parton energy to zero if $\Delta E\geq E$.
\end{Itemize}

The resulting energy loss is larger in the {\it non-reweighted} case, 
where partons are `absorbed' by the medium with a probability
$\int_E^\infty\d\epsilon\,P(\epsilon)$.
As we will see in the next section, 
the difference can be quite large for low $\pt$ 
and sufficiently-large transport coefficients.   
It is argued~\cite{carlosurs,wiedemann2} that
the difference in the values of the observables for the two approaches 
illustrates the theoretical uncertainties.
Along the lines of what is done in 
a recent work~\cite{heli} developed in parallel
to the present study, we display our model results as a band delimited by 
a solid line representing the {\it non-reweighted} case (larger quenching)
and a dashed line representing the {\it reweighted} case (smaller quenching). 
For the time being, from the theory side both approaches are highly 
disputable, while the guidance given by the experimental results will be 
commented in the conclusions.

%% file: src/results.tex
\section{Results}
\label{results}

\subsubsection*{Nuclear modification factor in \mbox{Au--Au} at 
$\sqrtsNN=200~\gev$}

\begin{figure}[t!]
  \begin{center}
    \includegraphics[width=0.49\textwidth]{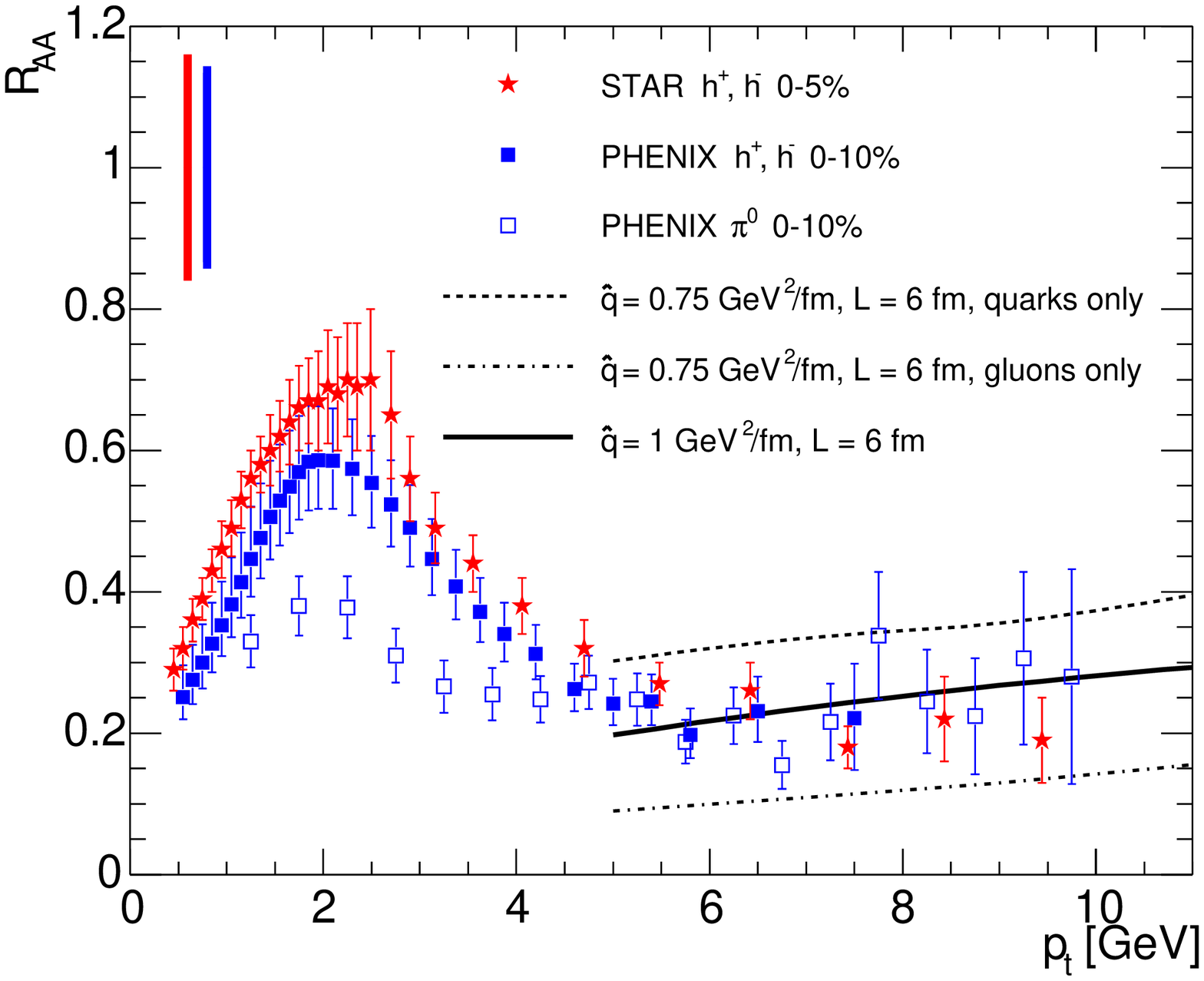}
    \includegraphics[width=0.49\textwidth]{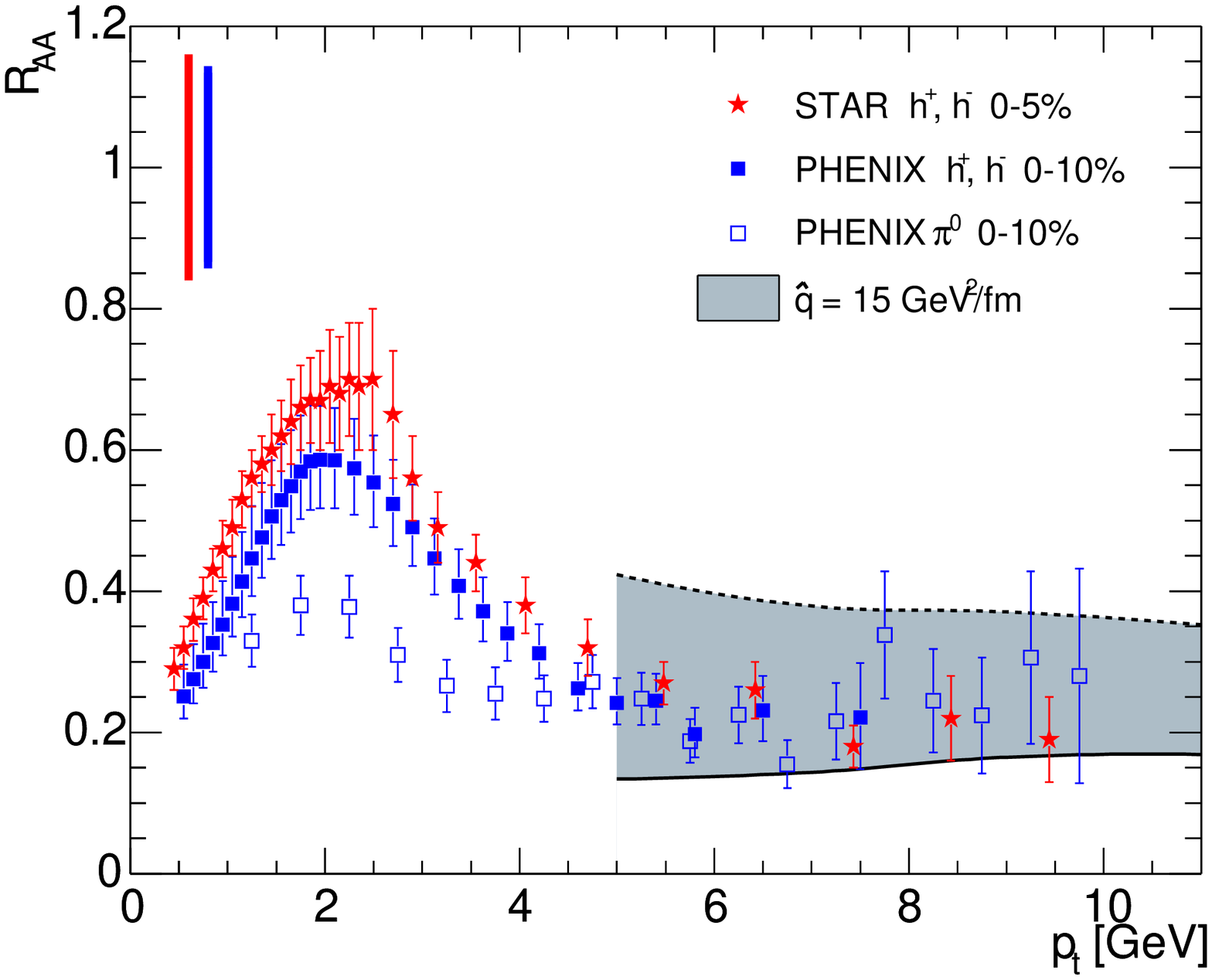}
    \caption{$\RAA(\pt)$ for central Au--Au collisions at 
             $\sqrtsNN=200~\gev$. PHENIX~\cite{phenixRAA} 
             and STAR~\cite{starRAA} data are reported with combined
             statistical and $\pt$-dependent systematic errors (bars on the 
             data points) and $\pt$-independent systematic errors (bars at 
             $\RAA=1$). Model results for constant $\hat{q}$ and $L$ 
             (left-hand panel) and for constant $\hat{q}$ and Glauber-based
             $L$ distribution (right-hand panel) are reported.
             In the right-hand plot and in all the following figures
             the shaded band is delimited by {\it non-reweighted} case (solid
             line) and {\it reweighted} case (dashed line).} 
    \label{fig:RAAq}
  \end{center}
\end{figure}

We start by presenting the results on $R_{\rm AA}(\pt)$ in central 
Au--Au collisions at $\sqrtsNN=200~\gev$ obtained using 
constant in-medium path length and transport coefficient (left-hand panel of 
Fig.~\ref{fig:RAAq}).  
The data on charged hadrons and neutral pions from 
PHENIX~\cite{phenixRAA} and STAR~\cite{starRAA}
are reported with combined statistical and $\pt$-dependent systematic errors
shown by the bars on the data points 
and $\pt$-independent normalization errors 
shown by the bars centred at $R_{\rm AA}=1$. The model results are shown 
by the lines: for all hadrons, with $\hat{q}=1~\gev^2/\fm$ and $L=6~\fm$,
(solid line) and 
for hadrons coming from quarks and from gluons, separately, with 
$\hat{q}=0.75~\gev^2/\fm$ and $L=6~\fm$, (dashed and dot-dashed lines). 
In order to compare our results to those in Ref.~\cite{carlosurs},
we use the same parameters and treat the finite-energy constraint in 
the {\it non-reweighted} case.
The two lines obtained with $\hat{q}=0.75~\gev^2/\fm$ and $L=6~\fm$
agree with those reported in Fig.~20 of Ref.~\cite{carlosurs}.
Since the high-$\pt$ hadron spectrum at RHIC energies is mainly coming from 
quarks\footnote{At $\sqrt{s}=200~\gev$, with CTEQ\,4L 
parton distribution functions~\cite{cteq4}, gluons dominate the 
parton $\pt$ distribution up to about $20~\gev$. However, since quarks 
fragment harder than gluons, high-$\pt$ hadrons are mostly produced
from quark fragmentation. Using 
KKP fragmentation functions~\cite{kkp}, we find that  
75\% of the hadrons with $\pt>5~\gev$ 
come from quark fragmentation and 25\% from gluon fragmentation.}, 
which lose less energy than gluons, a larger $\hat{q}$ of 
$\simeq 1~\gev^2/\fm$ is necessary to match the measured $\RAA$, 
when a realistic quarks-to-gluons ratio is used.

Before moving to the parton-by-parton approach of PQM outlined in the previous 
sections, it is very instructive to show the model results obtained using a 
constant transport coefficient and the Glauber-based path-length distribution 
for 0--10\% central collisions (displayed in the bottom-left panel of 
Fig.~\ref{fig:geo}). We have to use $\hat{q}\simeq 15~\gev^2/\fm$ 
to describe the data with the model band delimited by the {\it reweighted} and 
{\it non-reweighted} cases (right-hand panel of Fig.~\ref{fig:RAAq}).
When going from a constant $L=6~\fm$ to a realistic distribution, 
the transport coefficient has to be increased by more than one order of 
magnitude, because there 
are many partons with small path lengths of $2$--$3~\fm$ that can be quenched 
only if the medium is very dense. It is interesting to note that $\RAA$ 
is clearly increasing with $\pt$ when a constant length is used, while it is 
flatter with the full distribution. This is due to the presence of a 
long tail in the $L$ distribution, up to $12~\fm$: 
only high-energy partons can fully 
`exploit' this tail, while low-energy ones are just completely stopped by 
the medium after a few fm, so that the `effective' average length 
increases with the parton energy. 
We note that between the {\it non-reweighted}
and {\it reweighted} approach to the parton finite-energy constraint
there is a difference of about a factor 2 in the magnitude of $\RAA$,
but also a difference in the slope versus $\pt$, which is slightly positive 
for {\it non-reweighted} and slightly negative for {\it reweighted}.

Using a constant transport coefficient of $15~\gev^2/\fm$ 
and a realistic $L$ distribution, the measured 
hadron suppression can be fairly well described for $\pt \gsim 5~\gev$
(at lower $\pt$ we do not apply the model, as initial-state effects 
and in-medium hadronization, that we do not include, 
might play an important role). Remarkably, our result 
agrees with that obtained in Ref.~\cite{heli}, where the same quenching 
weights and a simplified collision geometry with effective nuclei  
(cylindrical density profile instead of the Wood-Saxon we use) 
are coupled to a leading-order perturbative QCD calculation. The $\RAA$ 
band is found to have similar $\pt$-dependence (rather flat) and width.
Numerically, the extracted value 
of $\hat{q}$ is $\simeq 10~\gev^2/\fm$ in Ref.~\cite{heli}, smaller than 
our $15~\gev^2/\fm$. However, this is not an inconsistency, since 
the value of $\alphas$ used in the calculation of the quenching weights
is 1/2 in Ref.~\cite{heli} and 1/3 here, and the scale
of the energy loss is set by the product $\alphas\,\hat{q}$ 
(see Eq.~(\ref{eq:avdE})).

\begin{figure}[t!]
  \begin{center}
    \includegraphics[width=0.9\textwidth]{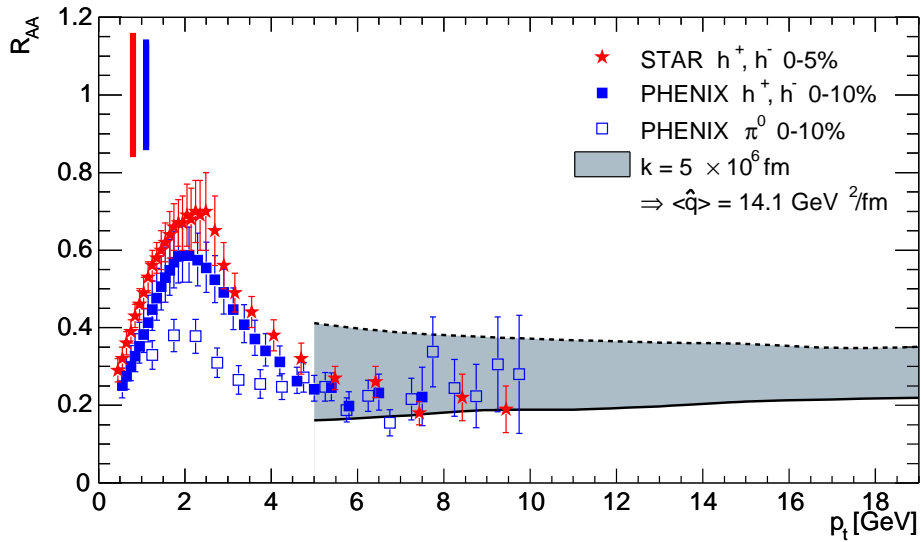}
    \caption{$\RAA(\pt)$ for central Au--Au collisions 
             at $\sqrtsNN=200~\gev$. The model band is obtained with 
             a parton-by-parton calculation of $\omegac$ and $R$. 
             The average transport coefficient is $14~\gev^2/\fm$.} 
    \label{fig:RAAkCentral}
  \end{center}
\end{figure}

\begin{figure}[!t]
  \begin{center}
    \includegraphics[width=0.49\textwidth]{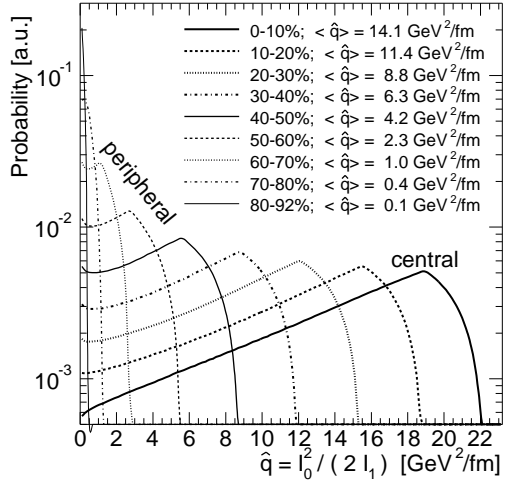}
    \caption{Distributions of $\hat{q}$, calculated from Eq.~(\ref{eq:L}), 
             for different centralities; 
             the $k$ parameter is fixed to the value that allows to describe
             $\RAA$ for the most central collisions.}
    \label{fig:QAllCentralities}
  \end{center}
\end{figure}

\begin{figure}[!t]
  \begin{center}
    \includegraphics[width=0.83\textwidth]{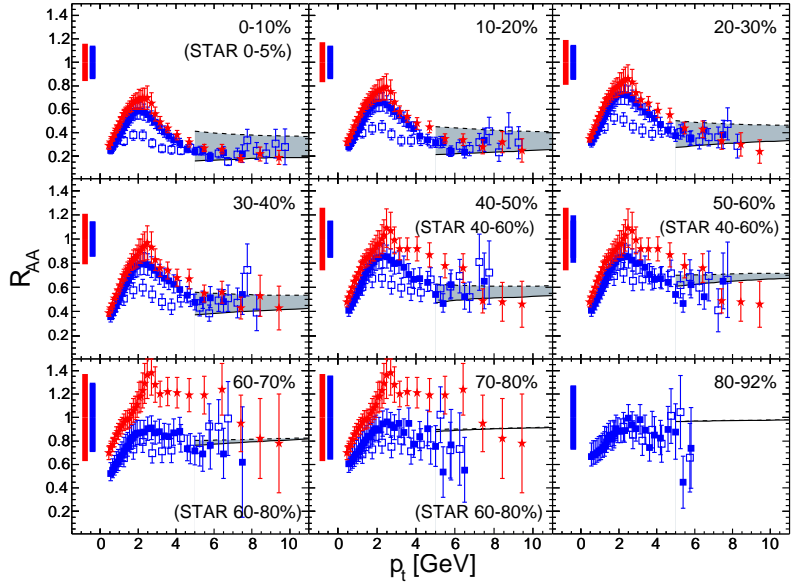}
    \caption{$\RAA(\pt)$ for different centralities. Data are PHENIX
             charged hadrons (closed squares) and $\pi^0$ 
             (open squares)~\cite{phenixRAA}
             and STAR charged hadrons (stars)~\cite{starRAA}.} 
   \label{fig:RAAkAllCentralities}
  \end{center}
\end{figure}

In order to address the centrality dependence of the high-$\pt$ suppression, 
we move to the parton-by-parton approach. For central collisions, the 
result obtained with the scale parameter $k=5\times 10^6~\fm$, 
corresponding to $\av{\hat{q}}\simeq 14~\gev^2/\fm$,
is shown in Fig.~\ref{fig:RAAkCentral}. The model band 
is very similar to that reported in the right-hand panel of Fig.~\ref{fig:RAAq}
for $\hat{q}=15~\gev^2/\fm$ and the $L$ distribution. 
We now vary the centrality, keeping always the same scale $k$. 
Figure~\ref{fig:QAllCentralities} shows the distributions
of $\hat{q}$, calculated from Eq.~(\ref{eq:L}), for different centrality bins.
The $\hat{q}$ variation within a given bin reflects the different 
parton production points, hence different medium densities encountered.
The rightmost (highest) 
value refers to partons originating in the centre of the 
collision system.
The model nuclear modification 
factors, compared to PHENIX~\cite{phenixRAA} and STAR~\cite{starRAA} data
ranging in centrality from 0--5\% to 80--92\%, are reported in 
Fig.~\ref{fig:RAAkAllCentralities}.
We note that the theoretical uncertainty band is narrower for 
semi-central and peripheral collisions, where, due to 
the smaller size and density of the medium, 
the probability to have $\Delta E>E$ in the quenching weights becomes 
marginal.

\begin{figure}[!t]
  \begin{center}
    \includegraphics[width=0.49\textwidth]{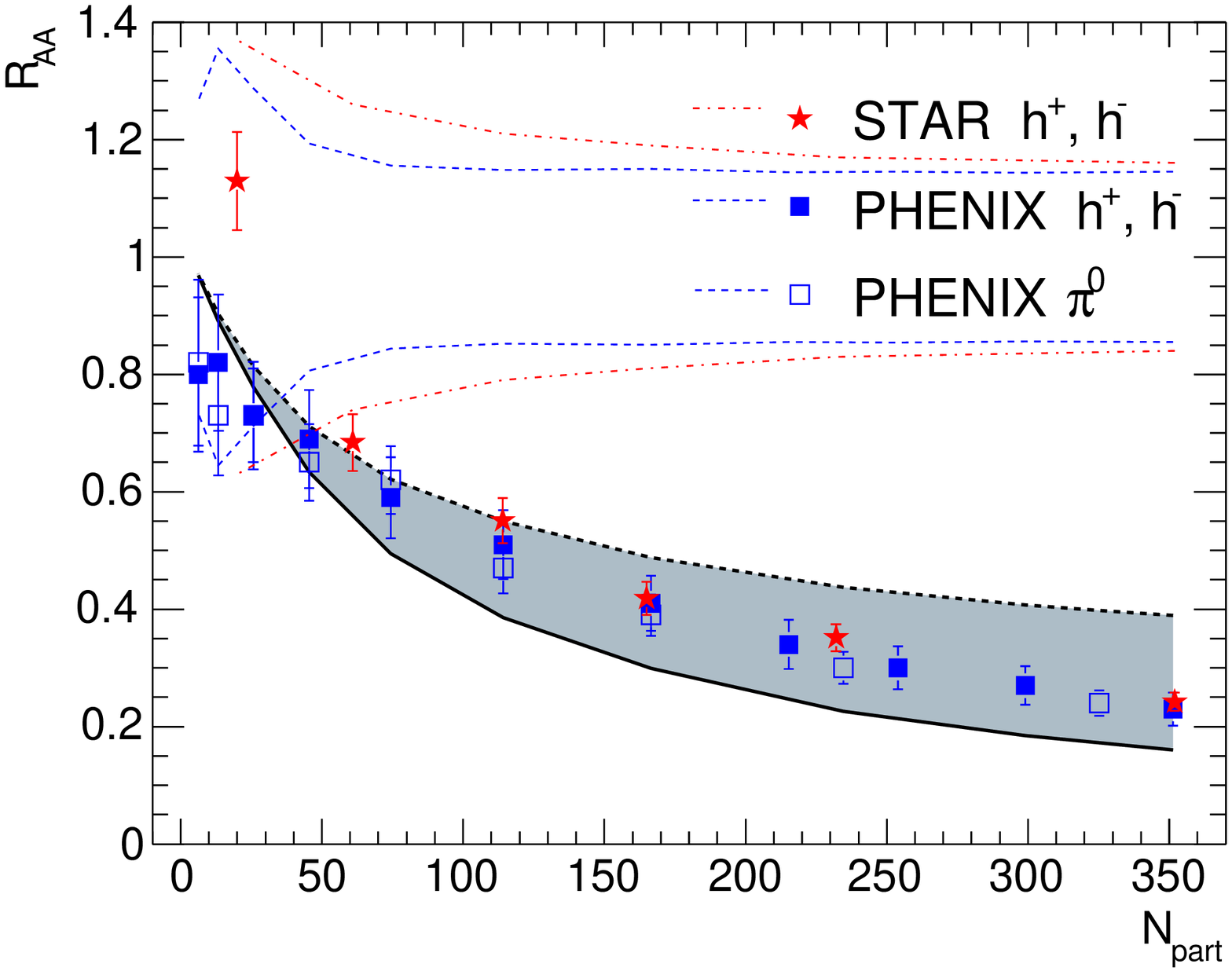}
    \includegraphics[width=0.49\textwidth]{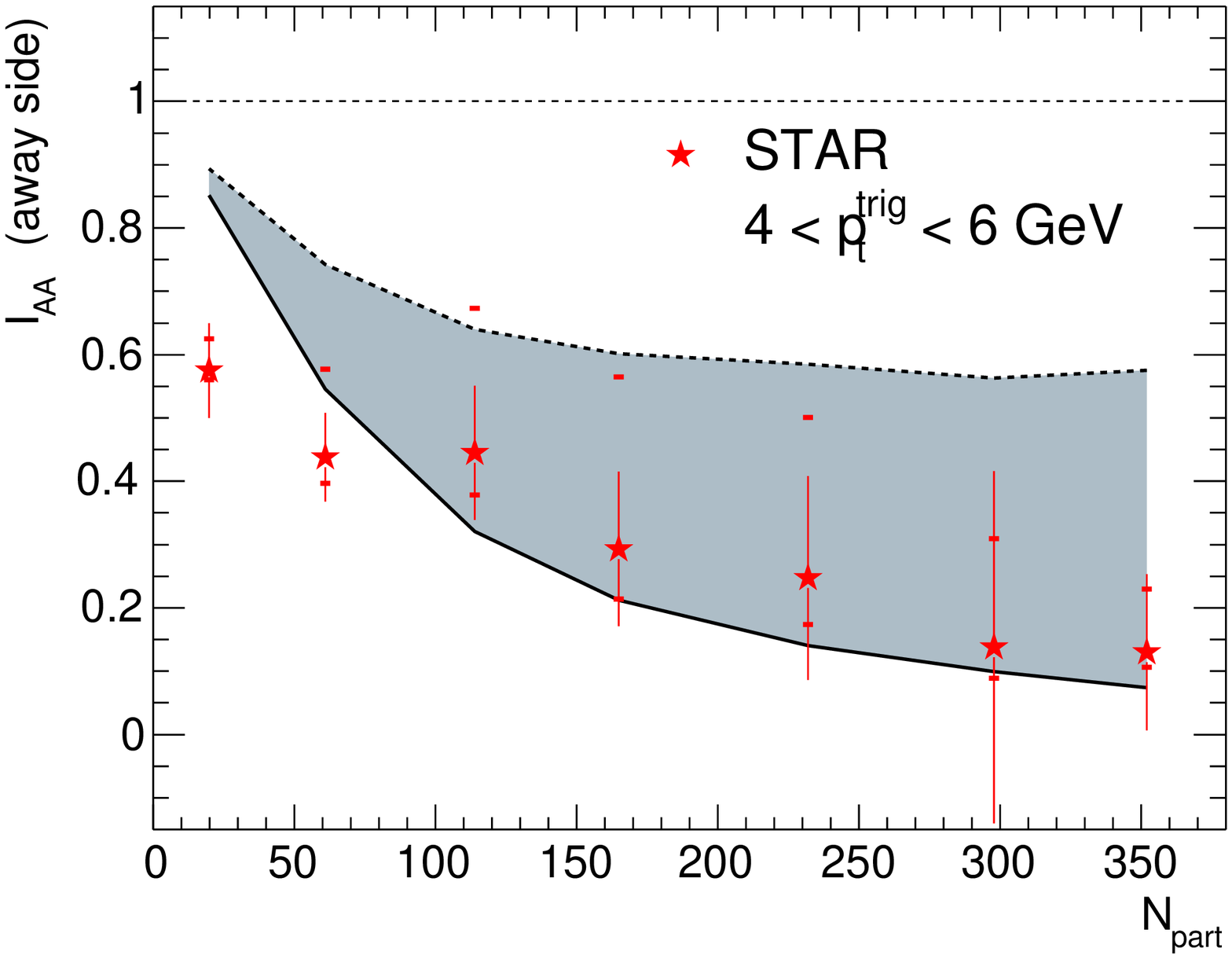}
    \caption{Average $\RAA$ in the range 
             $4.5<\pt<10~\gev$~\cite{phenixRAA,starRAA} (left-hand panel)
             and $I_{\rm AA}$, defined in the text, 
             for the away-side jet~\cite{starIAA} (right-hand panel)
             as a function of collision centrality, expressed by 
             the number of participants, $N_{\rm part}$. 
             For $\RAA$, the error bars are combined statistical and 
             $\pt$-dependent systematic errors and the bands 
             centred at $\RAA=1$ are the $\pt$-independent normalization 
             errors for PHENIX
             (dashed) and STAR (dot-dashed). 
             For $I_{\rm AA}$, the statistical
             (bars) and systematic (ticks) errors are shown.}
    \label{fig:RAAIAAvsNpart}
  \end{center}
\end{figure}

Our results follow the decrease of the measured $\RAA$ with increasing 
centrality. This is 
more conveniently visualized in the left-hand panel of 
Fig.~\ref{fig:RAAIAAvsNpart},
where we show the average $\RAA$ in the range $4.5<\pt<10~\gev$ plotted as 
a function of the number of participant nucleons, $N_{\rm part}$, 
obtained from the 
Glauber model. Data are taken from Refs.~\cite{phenixRAA,starRAA}.

\subsubsection*{Back-to-back correlations}

By generating pairs of back-to-back partons, 
we can study the centrality dependence of the disappearance of the
away-side jet. 
This effect is usually quantified using the correlation 
strength~\cite{wangDAA}
\begin{equation}
\label{eq:daa}
D_{\rm AA} =
\int^{p_{\rm t,1}}_{\pt^{\rm min}}\d p_{\rm t,2}
\int_{\Delta\phi>\Delta\phi^{\rm min}}\d\Delta\phi\,
\frac{\d^3\sigma_{\rm AA}^{h_1h_2}/\d p_{\rm t,1}\d p_{\rm t,2}
\d\Delta\phi}
{\d\sigma_{\rm AA}^{h_1}/\d p_{\rm t,1}}
\end{equation} 
for an associated hadron $h_2$ with transverse momentum 
$p_{\rm t,2}$ in the opposite azimuthal
direction of a trigger hadron $h_1$ with transverse momentum $p_{\rm t,1}$. 
The STAR data~\cite{starIAA} 
are for trigger particles with $4<p_{\rm t,1}<6~\gev$ 
and associated particles with $p_{\rm t,2}>\pt^{\rm min} = 2~\gev$ and 
$p_{\rm t,2}<p_{\rm t,1}$, with 
$\Delta\phi\equiv |\phi_1-\phi_2|>\Delta\phi^{\rm min}=130^\circ$. 
The correlation strength is then corrected for combinatorial
background and azimuthal anisotropy of particle production in 
non-central collisions~\cite{starIAA}. The correlation strength 
in nucleus--nucleus relative to pp collisions defines the suppression factor:
\begin{equation}
\label{eq:iaa}
I_{\rm AA}=\frac{D_{\rm AA}}{D_{\rm pp}}\,.
\end{equation}
We generate pairs of partons with the same initial $\pt$ and 
separated in azimuth by $\Delta\phi=180^\circ$. 
Then, we calculate $\omegac$ and $R$
for each parton and apply energy loss and fragmentation. We count as 
trigger particle every hadron $h_1$ with $4<p_{\rm t,1}<6~\gev$ 
and as associated away-side particle the other hadron $h_2$ of the pair, if 
its transverse momentum is in the range 
$2~\gev<p_{\rm t,2}<p_{\rm t,1}$. We define:
\begin{equation}
\label{eq:iaaus}
I_{\rm AA}=\left(\frac{N^{\rm associated}}{N^{\rm trigger}}\right)_{\rm with~energy~loss} \bigg/ \left(\frac{N^{\rm associated}}{N^{\rm trigger}}\right)_{\rm w/o~energy~loss}\,.
\end{equation}

The right-hand panel of Fig.~\ref{fig:RAAIAAvsNpart} shows our result
for $I_{\rm AA}$ versus $N_{\rm part}$,
compared to STAR measurements in Au--Au collisions at $\sqrtsNN=200~\gev$, 
with statistical (bars) and systematic (ticks) errors, 
from Ref.~\cite{starIAA}. The magnitude and centrality dependence of the 
suppression are described without changing the scale parameter value we
extracted from $\RAA$ in central collisions. 

\subsubsection*{Azimuthally-differential observables}

For non-central collisions, the nucleus--nucleus overlap profile is asymmetric
with respect to the event-plane direction, defined by the line that contains 
the centres of the two colliding nuclei and the impact parameter vector 
$\vec{b}$, in the transverse plane. The asymmetry 
is visible in the upper row of 
Fig.~\ref{fig:geo}, where the event-plane direction is parallel to the 
$x$ axis. Consequently, the in-medium path length is, 
on average, larger for partons propagating in the out-of-plane direction 
(perpendicular to the event plane) than for partons propagating in the 
in-plane direction (parallel to the event plane).

\begin{figure}[!t]
  \begin{center}
    \includegraphics[width=0.49\textwidth]{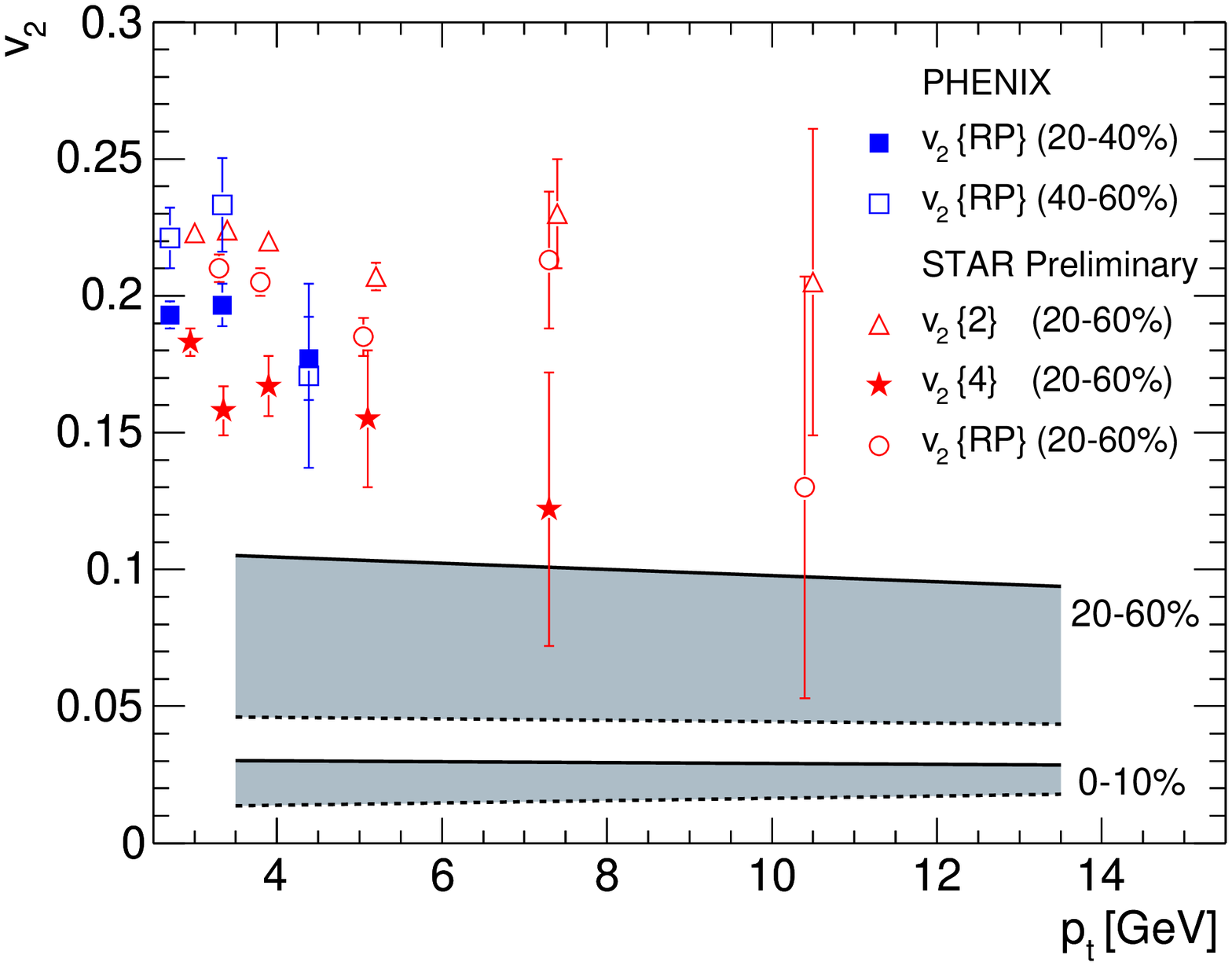}
    \includegraphics[width=0.49\textwidth]{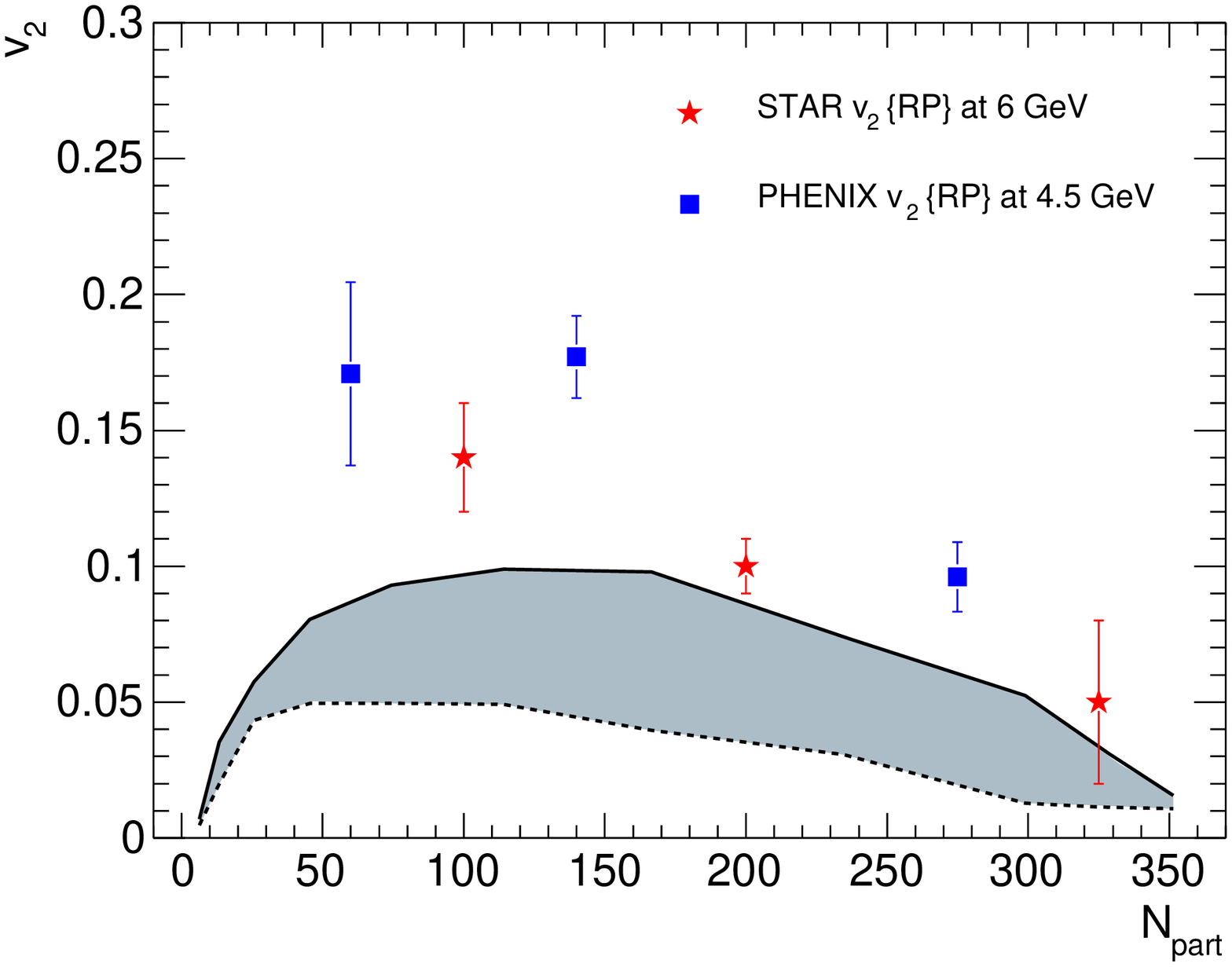}
    \caption{Transverse momentum and centrality dependence of the 
             azimuthal anisotropy $v_2$, compared to measurements for 
             charged hadrons from PHENIX~\cite{v2phenix} and
             STAR~\cite{v2starQM04,v2starFilimonov}
             (see text). Only the statistical errors
             are plotted. Note that, opposed to all other figures, here 
             the model result for the {\it non-reweighted} case (solid) is
             the upper limit of the band and that for the {\it reweighted}
             case (dashed) is the lower one.}
    \label{fig:v2}
  \end{center}
\end{figure}
 
Due to parton energy loss, the asymmetry in the medium geometry 
should be reflected in the azimuthal distribution $\d N/\d\phi$ 
of high-$\pt$ hadrons with respect to the event plane, $\phi=0^\circ$. 
We quantify this effect by calculating:
\begin{Itemize}
\item the azimuthal anisotropy, as given by the second Fourier 
      coefficient of the $\d N/\d\phi$ distribution, $v_2$~\cite{v2th};
      we obtain the value of $v_2$ for hadrons in a given 
      $\pt$ range by fitting their 
      azimuthal distribution to the form $a\cdot(1+2\,v_2\,\cos 2\phi)$;
\item $\RAA^{\phi_0}(\pt)$, the nuclear modification factor for hadrons 
      in an azimuthal cone of $45^\circ$ 
      centred at the angle $\phi_0$ with respect to the event plane;
      we use $\phi_0=0^\circ$ (in-plane), $\phi_0=90^\circ$ (out-of-plane) 
      and $\phi_0=45^\circ$ (intermediate);
\item $I_{\rm AA}^{\phi_0}$~(away side), the nucleus--nucleus away-side 
      correlation strength relative to pp, in the three azimuthal regions 
      defined for $\RAA^{\phi_0}$.  
\end{Itemize}
The scale parameter $k$ is again kept to the value that allows to 
match the measured $\RAA$ in central collisions at $\sqrtsNN=200~\gev$.

Figure~\ref{fig:v2} (left-hand panel) shows the model results for $v_2$ as a 
function of the transverse momentum, for central (0--10\%, 
$N_{\rm part}\approx 320$) and 
non-central (20--60\%, $N_{\rm part}\approx 100$) 
Au--Au collisions, compared to non-central experimental 
measurements on charged hadrons 
obtained by PHENIX~\cite{v2phenix} and STAR (preliminary)~\cite{v2starQM04}  
using three different methods: 
reaction plane reconstruction ($v_2\,\{{\rm RP}\}$),
2-particle correlations ($v_2\,\{2\}$)~\cite{v2methodsstar} and 
4-particle correlations ($v_2\,\{4\}$)~\cite{v2methodsstar}.
In the right-hand panel of the same figure, the $v_2$ centrality dependence 
from the model is compared to charged hadrons data from 
PHENIX~\cite{v2phenix}, 
at $\pt\approx 4.5~\gev$, and from STAR (preliminary)~\cite{v2starFilimonov}, 
at $\pt\approx 6~\gev$. 

The measured azimuthal anisotropy at intermediate 
transverse momenta of $4$--$6~\gev$ is systematically larger than that 
generated by parton energy loss in our model, indicating the presence of 
non-negligible collective flow effects in this momentum 
range. However, the preliminary STAR measurements at higher $\pt$, 
shown in the left-hand panel of Fig.~\ref{fig:v2}, suggest that 
$v_2$ might go down to values compatible with those expected from 
parton energy loss in an azimuthally-asymmetric medium.
High-$\pt$ data with larger statistics from the recent RHIC Run-4 will
allow to clarify this point. 
We note that our maximum $v_2$ of $0.05$--$0.10$, 
for $N_{\rm part}\approx 100$, is similar to that obtained in other 
parton energy loss~\cite{wangDAA} or absorption~\cite{drees} calculations. 

\begin{figure}[!t]
  \begin{center}
    \includegraphics[width=0.49\textwidth]{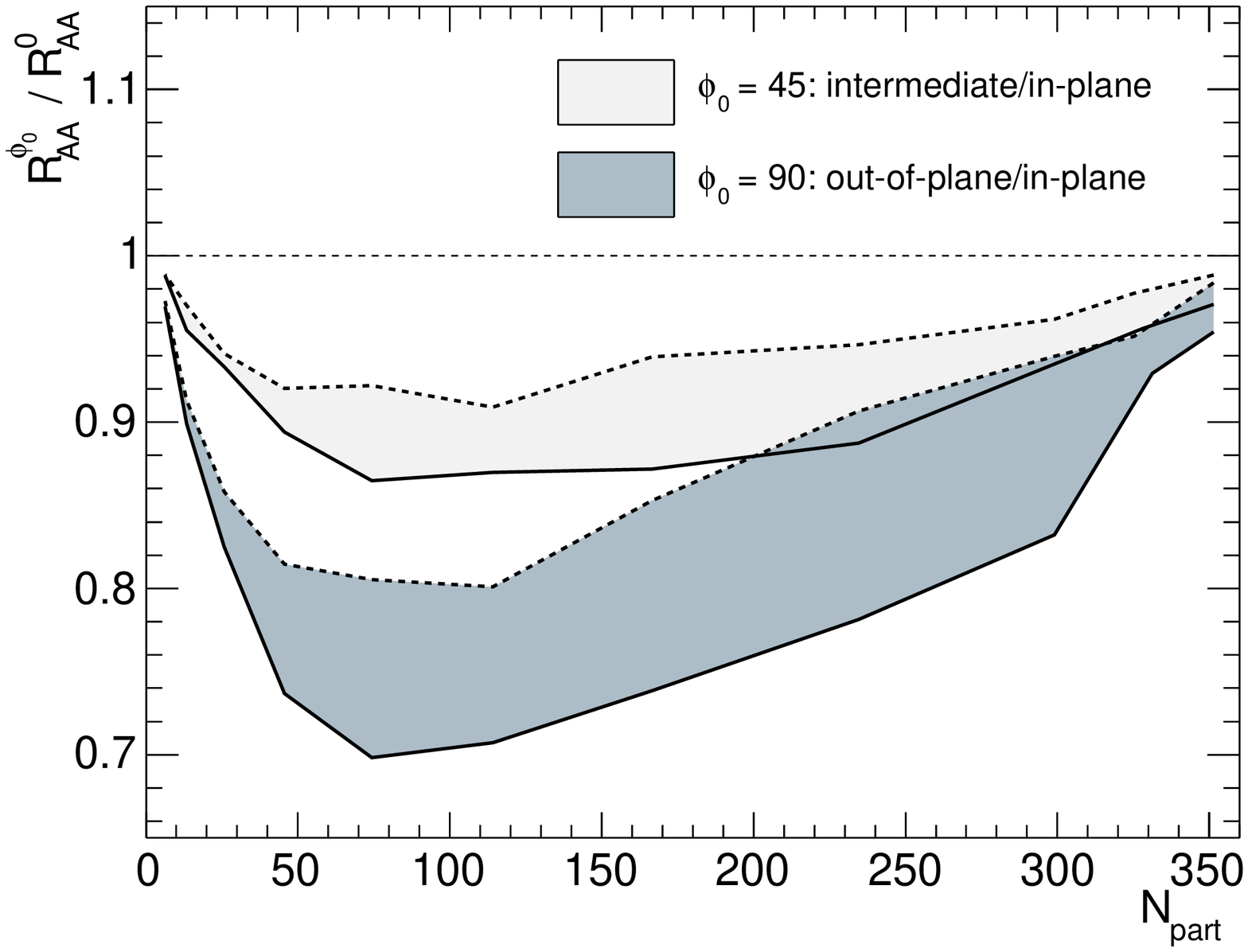}
    \includegraphics[width=0.49\textwidth]{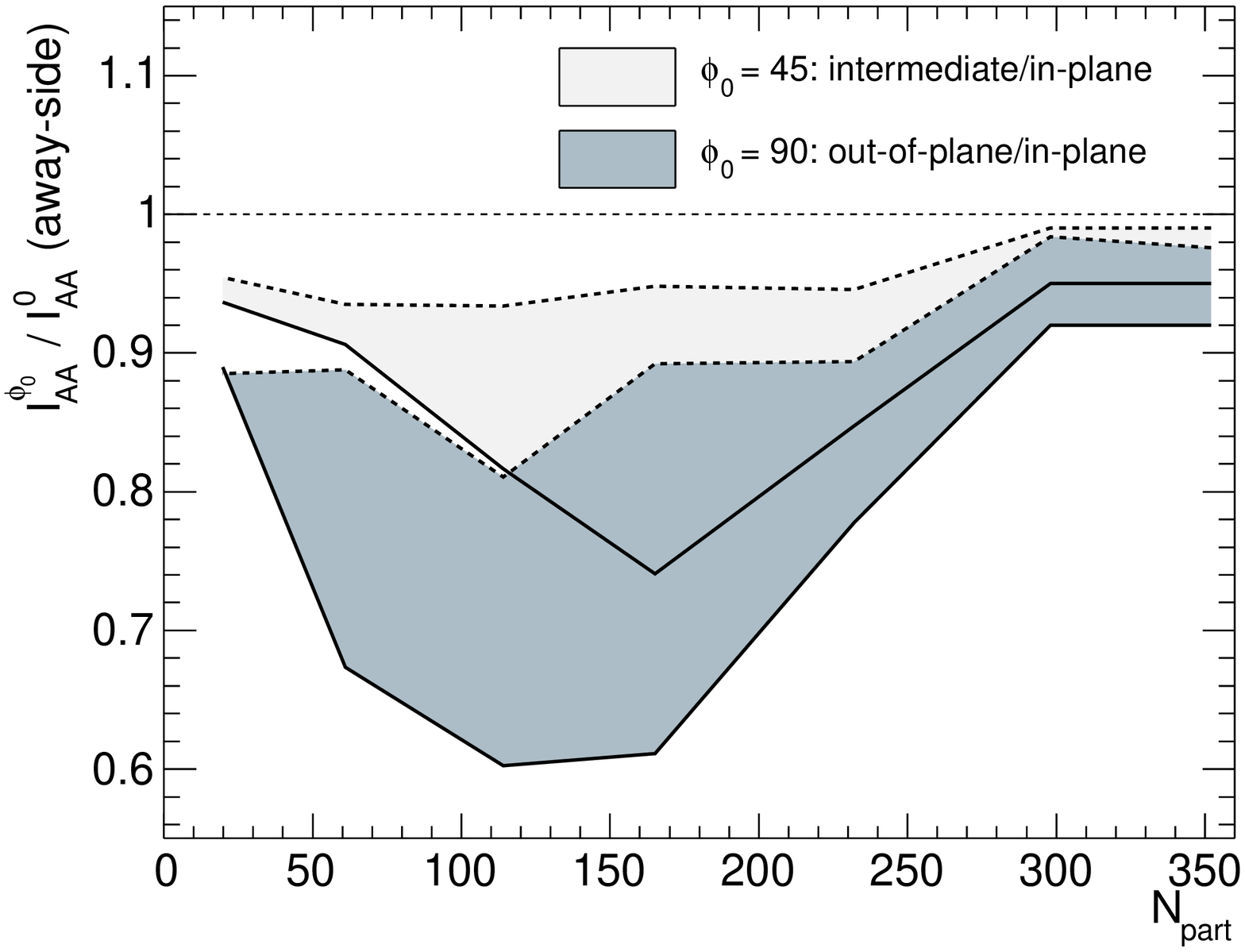}
    \caption{Azimuthal variation of the nuclear modification factor
             of the away-side correlations. In the left-hand panel, ratios
             $\RAA^{90}/\RAA^0$ (out-of-plane/in-plane) and 
             $\RAA^{45}/\RAA^0$ (intermediate/in-plane), averaged over the 
             range $4.5<\pt<10~\gev$.
             In the right-hand panel, the same ratios for $I_{\rm AA}$ 
             with trigger conditions as in Fig.~\ref{fig:RAAIAAvsNpart}.
             Both observables are plotted as a function of $N_{\rm part}$.}
    \label{fig:RAAIAAphiRHIC}
  \end{center}
\end{figure}

The azimuthal variation of the nuclear modification factor and of 
the away-side correlations is illustrated in Fig.~\ref{fig:RAAIAAphiRHIC}. 
For $\RAA$ (left-hand panel), we show the two ratios
$\RAA^{90}/\RAA^0$ (out-of-plane/in-plane) and $\RAA^{45}/\RAA^0$
(intermediate/in-plane),
averaged over the range $4.5<\pt<10~\gev$,  
as a function of collision centrality ($N_{\rm part}$).
As for $v_2$, the asymmetry is maximum at $N_{\rm part}\approx 100$,
where the model gives for $\RAA$ a ratio out-of-plane/in-plane of 
$\approx 0.75$. 
Similarly, for the away-side correlation $I_{\rm AA}$ (right-hand panel), 
we show the two ratios $I_{\rm AA}^{90}/I_{\rm AA}^0$ (out-of-plane/in-plane) 
and $I_{\rm AA}^{45}/I_{\rm AA}^0$ (intermediate/in-plane). The conditions
on the near-side trigger and the associated away-side
particles are the same as for Fig.~\ref{fig:RAAIAAvsNpart}.
At $N_{\rm part}\approx 100$--$150$ the model predicts 
an away-side correlation strength of about 30\% lower for the 
out-of-plane relative to the in-plane direction.
Both effects are rather strong and their measurement at RHIC 
would be of great interest.

\subsubsection*{Nuclear modification factor at $\sqrtsNN=62.4~\gev$}

The recent RHIC run with Au--Au collisions at $\sqrtsNN=62.4~\gev$ 
allows the measurement of the nuclear modification factor 
for charged hadrons and neutral pions up to transverse momenta 
of $7$--$8~\gev$. 
We estimate the leading-particle
suppression due to parton energy loss at this lower centre-of-mass energy
by using the proportionality of the transport coefficient $\hat{q}$ to 
the initial volume-density of gluons $n^{\rm gluons}$~\cite{baier}. 
In the saturation model~\cite{ekrt}, for collisions of two 
nuclei with mass number A at energy $\sqrtsNN$, 
such density is estimated to scale as 
\begin{equation}
\label{eq:ekrt}
n^{\rm gluons}\propto {\rm A}^{0.383}\, \left(\sqrtsNN\right)^{0.574}. 
\end{equation}
This gives $n^{\rm gluons}_{\rm Au-Au,\,62.4\,\gev}\simeq 0.5\times n^{\rm gluons}_{\rm Au-Au,\,200\,\gev}$. 
Applying this scaling to the value of the $k$ parameter, see 
Eq.~(\ref{eq:qofxi}), found in our model for central collisions 
at $200~\gev$, we obtain a transport coefficient distribution 
with mean value $\av{\hat{q}}\simeq 7~\gev^2/\fm$ in central collisions 
at $62.4~\gev$. 

\begin{figure}[!t]
  \begin{center}
    \includegraphics[width=.6\textwidth]{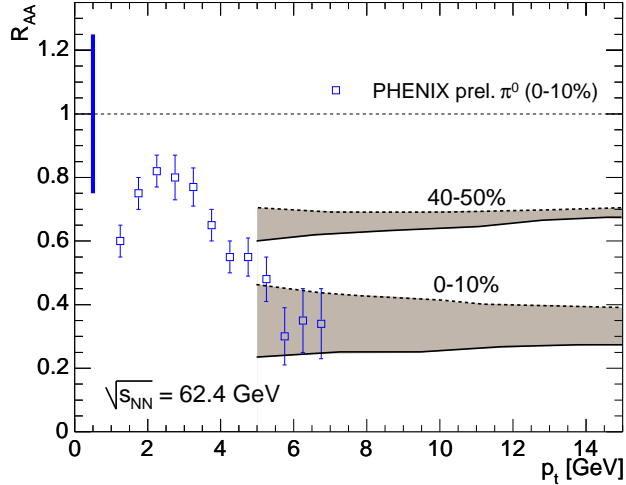}
    \caption{Model results for $\RAA(\pt)$ in central and semi-peripheral 
             Au--Au collisions at $\sqrtsNN=62.4~\gev$.
             The preliminary $\pi^0$ data (0--10\% centrality class) 
             from PHENIX~\cite{data62} are also shown; the pp reference
             is the PHENIX ${\rm pp}\to\pi^0+X$ parameterization,  
             the error bars on the data points are the combined 
             statistical and $\pt$-dependent systematic errors and the 
             bar centred at $\RAA=1$ is the systematic error on the 
             normalization.}
    \label{fig:RAA62}
  \end{center}
\end{figure}

We generate hard partons using PYTHIA at $\sqrt{s}=62.4~\gev$ and 
use the procedure described in Section~\ref{method}. The results 
are shown in Fig.~\ref{fig:RAA62}, along with preliminary data from 
PHENIX~\cite{data62} for neutral pions up to $\pt\approx 7~\gev$ 
in 0--10\% central collisions. For $\pt\gsim 5~\gev$, we find 
$\RAA\simeq 0.3$, in accordance with the data, in central (0--10\%) 
and $\simeq 0.7$ in semi-peripheral (40--50\%) collisions. 
These values are not much larger than those at $\sqrtsNN=200~\gev$. 
At smaller $\sqrtsNN$, although the transport coefficient is reduced by 
a factor of 2, the increased softness of the parton transverse momentum 
distribution determines a stronger effect of energy loss on the nuclear 
modification factor. Prior to the release of the preliminary PHENIX 
data, several predictions were published~\cite{vitev2004,wang2004,adil2004}. 
While the magnitude and $\pt$-dependence of $\RAA$ in Ref.~\cite{vitev2004} 
seem to agree with our result (although $\RAA$ is presented only up to 
$\pt=6~\gev$ there),
the predictions in Refs.~\cite{wang2004,adil2004} 
show a different trend with $\pt$: 
$\RAA$ for $\pt\gsim 5~\gev$ is decreasing with increasing $\pt$, down to
values of about 0.2 at $\pt\simeq 16~\gev$. 

\subsubsection*{Extrapolation to the LHC}
\label{lhc}

\begin{figure}[!t]
  \begin{center}
    \includegraphics[width=\textwidth]{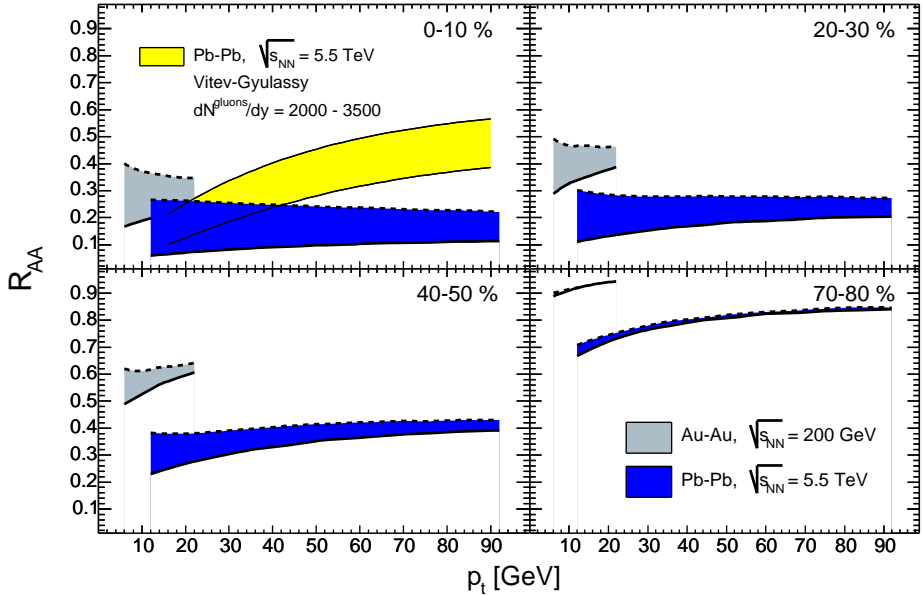}
    \caption{$\RAA(\pt)$ for different centrality classes in 
             Pb--Pb collisions at $\sqrtsNN=5.5~\tev$ and in Au--Au 
             collisions at $\sqrtsNN=200~\gev$. For comparison, the 
             prediction for LHC presented in Ref.~\cite{vitev}
             (Vitev--Gyulassy) is also shown.} 
    \label{fig:RAAlhc}
  \end{center}
\end{figure}

To compute the expected nuclear modification factor in \PbPb~collisions at 
the LHC we use PQM with the parton $\pt$ distribution extracted from PYTHIA 
at $\sqrt{s}=5.5~\tev$. Scaling 
the $k$ parameter according to Eq.~(\ref{eq:ekrt}), we have 
$n^{\rm gluons}_{\rm Pb-Pb,\,5.5\,\tev}\simeq 7\times n^{\rm gluons}_{\rm Au-Au,\,200\,\gev}$, 
i.e. $\av{\hat{q}}\simeq 100~\gev^2/\fm$. 

We report in 
Fig.~\ref{fig:RAAlhc} the expected transverse-momentum dependence of $\RAA$ in 
the range $10<\pt<90~\gev$ for different centralities (the results at 
$\sqrtsNN=200~\gev$ are shown as well).
In the most central collisions $\RAA$ is of $\approx 0.15$,
independent of $\pt$. This value is about a factor of 2 smaller 
than that measured at $\sqrtsNN=200~\gev$. 
Our result for the LHC is in agreement, both in the numerical value and in the 
$\pt$-dependence, with that obtained in Ref.~\cite{heli} using the 
same quenching weights and the same $\alphas\av{\hat{q}}$,
while it is quite different from that calculated in 
Ref.~\cite{vitev} assuming an initial gluon rapidity density 
$\d N^{\rm gluons}/\d y$ in the range 2000--3500.
For comparison, we have reported in the same figure the result of 
Ref.~\cite{vitev}:
there, $\RAA$ is predicted to rise significantly at 
large transverse momenta, from 0.1--0.2 at $20~\gev$ to 0.4--0.6 at 
$90~\gev$. 
We note that the difference between the two results 
is not likely to be due to the fact that we 
do not include nuclear (anti-)shadowing effects, since these are expected to 
determine a rather $\pt$-independent increase of $\RAA$ of about 10\% 
in the range $25<\pt<100~\gev$~\cite{kariheli,vitev}.

%% file: src/disc.tex
\section{Discussion}
\label{discussion}

\subsubsection*{High-energy partons from the surface}

The centrality dependence of leading-hadron suppression and back-to-back 
di-hadron correlations is well described by our model, in which the 
centrality evolution is purely given by collision geometry. This suggests 
that the high-opacity medium formed in \mbox{Au--Au} collisions 
at $\sqrtsNN=200~\gev$ has initial size and density that decrease from central 
to peripheral events according to the overlap profile of the colliding 
nuclei, $T_{\rm A}(x,y)\times T_{\rm B}(x,y)$. At the centre of the medium
the density is maximum and partons crossing this region are likely to 
be completely absorbed. Only partons produced in the vicinity of the surface
and propagating outward can escape from the medium with sufficiently-high 
energy to fragment into hadrons with more than few 
$\gev$ in $\pt$. Such an `emission from the surface' 
scenario was pictured also in a recent work~\cite{drees}, where the 
centrality dependence of $\RAA$ and $I_{\rm AA}$ could be reproduced by 
a simple model of parton absorption whose only physical ingredient was 
a Glauber-based nucleus--nucleus overlap profile. 

\begin{figure}[t!]
  \begin{center}
    \includegraphics[width=\textwidth]{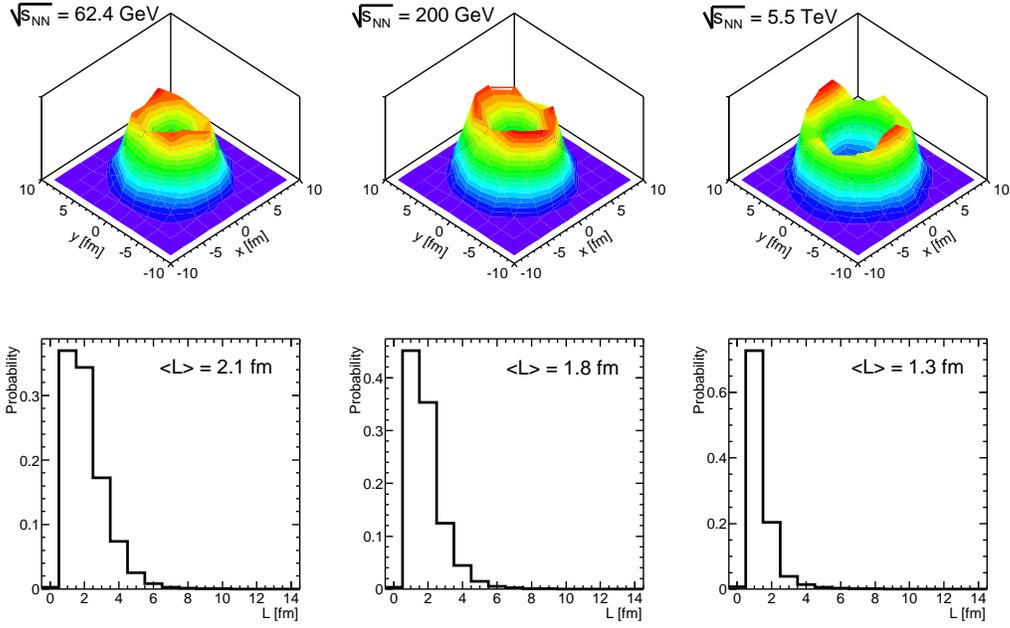}
    \caption{Distributions of parton production points in the transverse plane 
             (upper row)
             and in-medium path length (lower row) for partons that escape
             the medium and produce hadrons with $\pt>5~\gev$ 
             in central Au--Au collisions at 62.4 
             and 200~$\gev$ and in central Pb--Pb collisions at 5.5~$\tev$.
             The quantity $\av{L}$ is the average of the path length 
             distribution.
             These plots were obtained in the {\it non-reweighted} approach.} 
    \label{fig:survivors}
  \end{center}
\end{figure}
 
The region from which partons escape from the medium is visualized
by plotting the distribution of production points for partons that give 
a high-energy hadron ($\pt^{\rm hadron}>5~\gev$). This distribution 
for central Au--Au collisions at $62.4$ and $200~\gev$
and Pb--Pb collisions at $5.5~\tev$
is shown in Fig.~\ref{fig:survivors}, along with the corresponding path length 
distribution. The `thickness' of the escape region is of order 2--3~$\fm$ 
and it decreases as $\sqrtsNN$ increases from intermediate RHIC energy to 
LHC energy. 

\begin{figure}[!t]
  \begin{center}
    \includegraphics[width=0.45\textwidth]{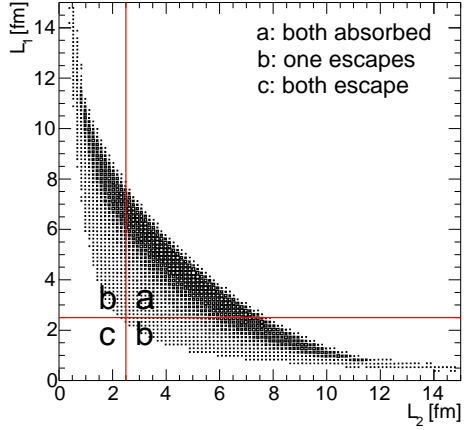}
    \caption{Correlation between path lengths for pairs of partons
             produced back-to-back in 0--10\% central Au--Au collisions.} 
    \label{fig:L1L2}
  \end{center}
\end{figure}

It is interesting to try to apply a simple toy model: all 
partons with a path length $L$ smaller than a maximum length 
$L^{\rm max}_{\rm escape}$ escape from the medium, the others are absorbed.
We define the path length probability distribution $\mathcal{P}(\ell)$ as the  
probability distribution for a generic parton to have a path length $\ell$. 
The distributions in the lower row of Fig.~\ref{fig:geo} are examples of 
$\mathcal{P}(\ell)$ for different centrality classes in \mbox{Au--Au}.
$\mathcal{P}(\ell)$ is normalized to unity, 
$\int_0^\infty \d\ell\,\mathcal{P}(\ell)=1$, and, thus, the integral 
$\int_0^L \d\ell\,\mathcal{P}(\ell)$ gives the fraction of partons with 
path length smaller than $L$. 
Using the measured (or expected) $\RAA$ for given collision energy and 
centrality, $L^{\rm max}_{\rm escape}$ can be estimated as 
$\int_0^{L^{\rm max}_{\rm escape}}\d\ell\,\mathcal{P}(\ell)=\RAA$.
At $\sqrtsNN=200~\gev$, we find $L^{\rm max}_{\rm escape}\approx 2.5~\fm$
from central (0--5\%) 
to semi-peripheral collisions (40--60\%): in this wide centrality 
range the simultaneous decrease of system density and volume results 
in energetic partons being emitted from a shell of constant thickness.
For more peripheral collisions the system becomes very diluted and 
partons can escape from the whole volume 
($L^{\rm max}_{\rm escape}\approx 3.5$--$4~\fm~\approx {\rm system~size}$).
In central collisions at different energies, 
we find $L^{\rm max}_{\rm escape}\approx 3~\fm$ at $\sqrtsNN=62.4~\gev$
and $L^{\rm max}_{\rm escape}\approx 1.5~\fm$ at $\sqrtsNN=5.5~\tev$ (LHC).

Remarkably, this absorption toy model allows to reconcile the magnitude 
of single-particle and away-side correlation suppressions measured in 
central Au--Au at 200~$\gev$, as illustrated in Fig.~\ref{fig:L1L2}.
The figure shows the distribution $L_1$ versus $L_2$ for pairs of partons 
(1 and 2) generated back-to-back. Parton pairs produced in the middle of 
the overlap profile populate the central part of the distribution 
($L_1 \sim L_2$), while pairs produced closer to the surface are in the 
two tails ($L_1 \gg L_2$ or $L_1 \ll L_2$). We report the two lines
$L_1=L^{\rm max}_{\rm escape}$ and $L_2=L^{\rm max}_{\rm escape}$, which 
divide the distribution in three parts: 
(a) for $L_{1,2}>L^{\rm max}_{\rm escape}$ both partons are absorbed,
(b) for $L_{1(2)}<L^{\rm max}_{\rm escape}$ and 
$L_{2(1)}>L^{\rm max}_{\rm escape}$ 
only one of the two partons escape the medium, and (c) for
$L_{1,2}<L^{\rm max}_{\rm escape}$ 
both partons escape. With the value $L^{\rm max}_{\rm escape}=2.5~\fm$,
extracted from the measured $\RAA$, the third part of the distribution (c)
is empty: it never happens that both partons can escape, in agreement 
with the value compatible with zero measured by STAR for $I_{\rm AA}$.

\subsubsection*{Energy-loss saturation}

\begin{figure}[t!]
  \begin{center}
    \includegraphics[width=\textwidth]{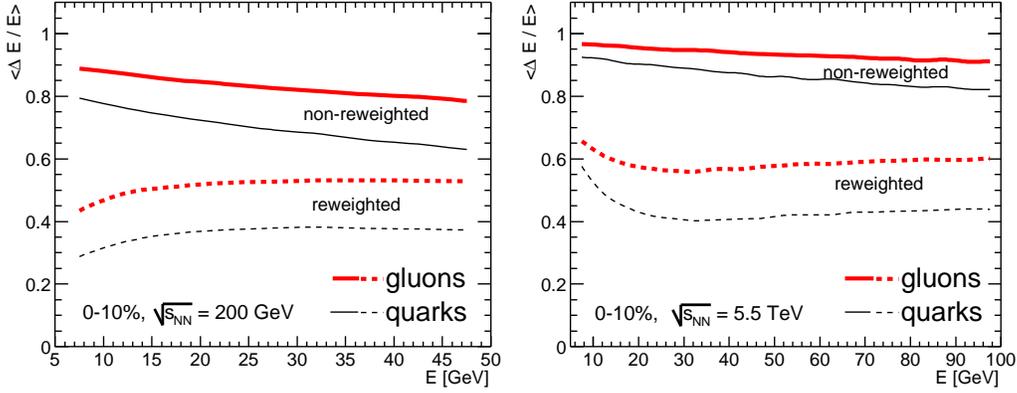}
    \caption{Average relative energy loss versus parton energy for
             quarks and gluons in central collisions at RHIC
             and LHC energies
             for the {\it non-reweighted} and {\it reweighted} cases.} 
    \label{fig:avgloss}
  \end{center}
\end{figure}

The strong parton absorption suggests that we are in a 
saturation regime of the energy loss, $\Delta E/E\to 1$, 
as almost all hard partons produced in the
inner core are thermalized ($\Delta E/E=1$) before escaping the medium. 
Indeed, the average relative energy loss, $\av{\Delta E/E}$
(from the Monte Carlo),
shown versus parton energy $E$ in Fig.~\ref{fig:avgloss} 
for central collisions at $\sqrtsNN = 200$ and $5500~\gev$,
is almost saturating to unity for gluons (70--80\%) and it is very 
large also for quarks (50--70\%). 
Due to the fact that gluons are closer
to energy-loss saturation than quarks, 
the ratio of gluon to quark $\av{\Delta E/E}$ 
is much smaller than the Casimir ratio $C_{\rm A}/C_{\rm F}=2.25$
expected from Eq.~(\ref{eq:avdE}). 
Furthermore, since absorption 
(and, hence, saturation) is more significant for small-$E$ partons, 
or, in other words, large-$E$ partons can exploit larger energy losses,
the genuine BDMPS $\Delta E/E\propto 1/E$ is replaced by 
a rather $E$-independent effective $\Delta E/E$. It is important to 
point out here that the $\pt$-independent nuclear modification factor 
obtained in our model, in agreement with RHIC data above $\approx 5~\gev$,
is a natural consequence of this saturation scenario. 

\begin{figure}[t!]
  \begin{center}
    \includegraphics[width=0.49\textwidth]{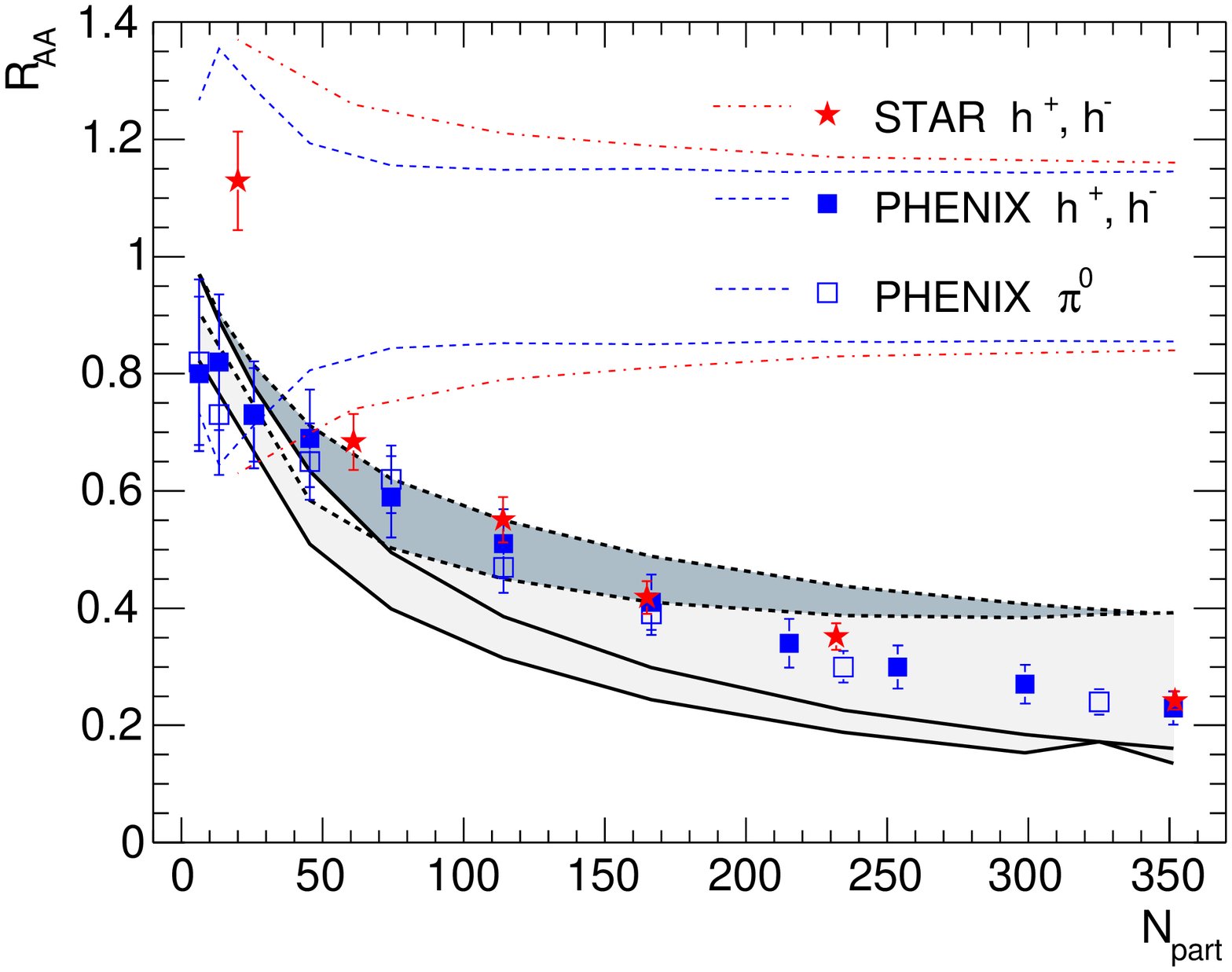}
    \includegraphics[width=0.49\textwidth]{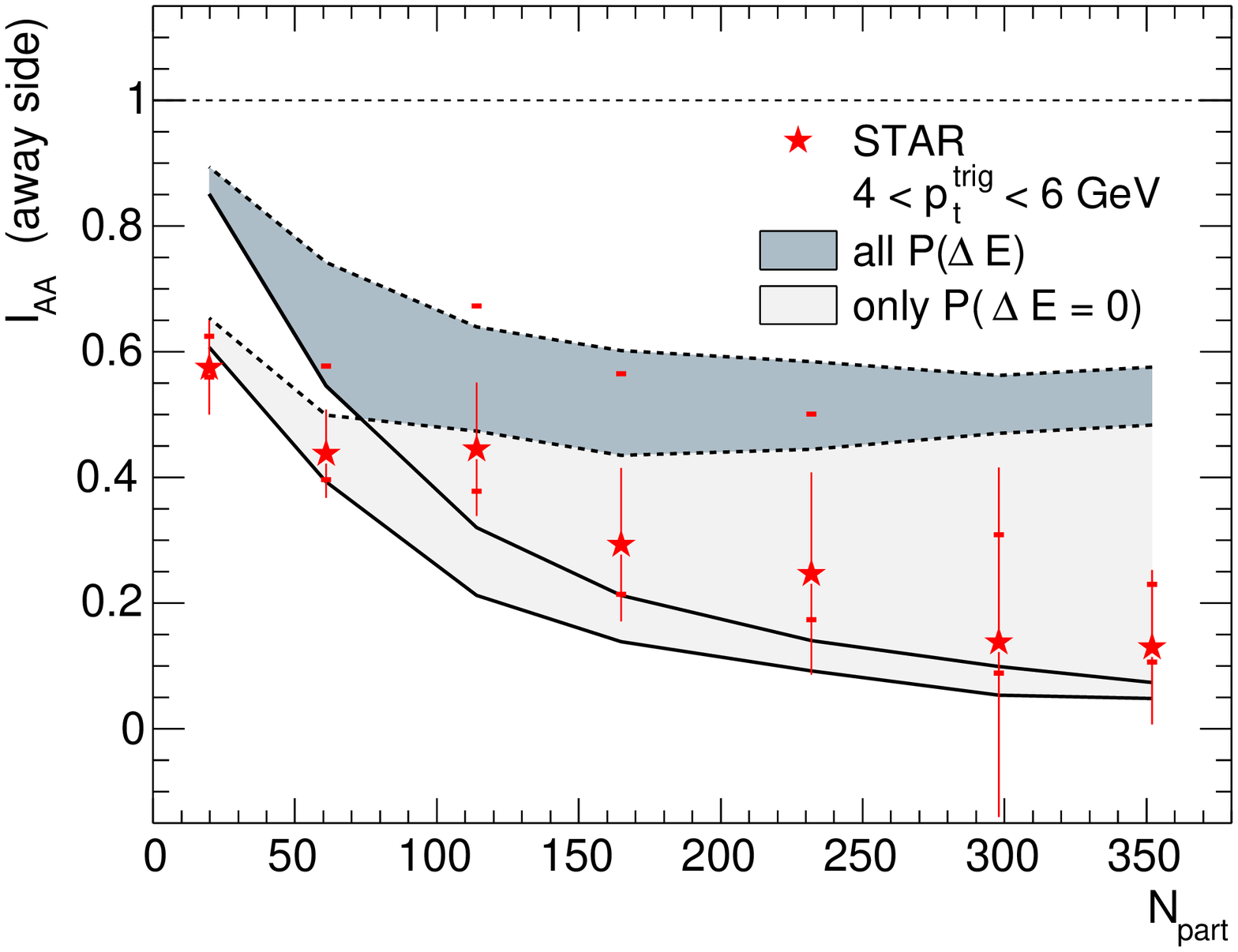}
    \caption{Average $\RAA$ in the range $4.5<\pt<10~\gev$ (left-hand panel)
             and $I_{\rm AA}$ for the away-side jet (right-hand panel)
             as a function of
             the number of participants. The data points and errors are
             the same as in Fig.~\ref{fig:RAAIAAvsNpart}.
             Here the model results obtained using either zero or maximum 
             energy loss, as explained in the text, (labelled 
             `only $P(\Delta E=0)$', lighter band) 
             are shown together with those 
             obtained with the standard procedure (labelled 
             `full $P(\Delta E)$', darker band).} 
    \label{fig:RAAIAAvsNpartP0only}
  \end{center}
\end{figure}

As the average relative energy loss is close to one, at least 
for the {\it non-reweighted} case, we are not very sensitive to the 
shape of the continuous part of the quenching weights in 
Eq.~(\ref{eq:pdeltae}), $p(\Delta E)$. 
Rather, the energy-loss probability is dominated by the discrete part, 
the probability to have no medium-induced radiation, $p_0$.
In order to confirm this statement, we repeat the calculation with
a modified PQM version: in the quenching procedure we
consider the parton as absorbed whenever the sampled energy loss $\Delta E$
is larger than zero. 
That is, we have either no energy loss or maximum energy loss. 
Also in this case, we consider the two finite-energy 
constraint methods, {\it non-reweighted} and {\it reweighted}.
We note that the `survival' probability is $p_0$ for the 
{\it non-reweighted} case and $p_0/\int_0^E \d\epsilon\,P(\epsilon)$
for the {\it reweighted} case.
In Fig.~\ref{fig:RAAIAAvsNpartP0only} we report, 
as a function of the number of participants, $R_{\rm AA}$ and
$I_{\rm AA}$, calculated with this modified quenching procedure,
labelled `only $P(\Delta E=0)$', and the same
value of $k$ we used for the standard procedure.
For the most central collisions, down to $N_{\rm part}\simeq 150$, 
the agreement with data is very good, whereas deviations are clearly
visible in $\RAA$ when going to semi-peripheral and peripheral 
collisions, $N_{\rm part}<150$. 
This confirms that, in central collisions, partons 
are either completely absorbed or coming from the surface, whereas in 
non-central collisions the shape of the energy-loss probability distribution
plays a role in the description of the data.

%% file: src/concl.tex
\section{Conclusions}
\label{conclusions}

Most of the present high-momentum observables have been studied 
using the Parton Quenching Model (PQM)
in which hard partons are generated with PYTHIA~\cite{pythia}, 
medium-modified with the quenching weights~\cite{carlosurs}, 
and hadronized independently via KKP fragmentation functions~\cite{kkp}.
Using a Glauber approach with Wood-Saxon density profiles of the colliding
nuclei, the chain takes into account
the realistic spatial distribution of hard parton production points 
and the amount and density of matter traversed by each parton.

The results show that, if parton production coordinates 
and realistic density profiles are taken into account,
the ensuing transport coefficient has to acquire very large values:
$\av{\hat{q}}\simeq 14~\gev^2/\fm$ in central \mbox{Au--Au} collisions
at $\sqrtsNN=200~\gev$. 
Note that we used a relatively small value of $\alphas$ (1/3). 
Due to the scaling $\Delta E\propto\alphas\,\hat{q}$, 
if larger values (e.g. 1/2) were 
used, the extracted transport coefficient would be smaller, 
but still quite large, $\av{\hat{q}}\simeq 10~\gev^2/\fm$~\cite{heli}. 
In Ref.~\cite{heli}, it is pointed out 
that such $\hat q$ values do not necessarily imply unexpectedly large
medium initial energy densities (a few hundreds $\gev/\fm^3$), 
as one obtains in the hypothesis of an ideal 
plasma whose constituents interact perturbatively with the hard 
partons~\cite{baier},
but rather suggest that the medium might interact with the hard partons 
much stronger than perturbatively expected. 
Technically, the large extracted $\hat q$
values present a yet unsolved theoretical problem, since
the eikonal approach used in the theory cannot be flawlessly extended to
finite (low) parton energies. In the situation, we presented
two possibilities:
\begin{Enumerate}
\item the theoretical treatment is applied regardless of the obvious problem
      for low-energy partons, which are completely absorbed with a rather 
      large probability ({\it non-reweighted} approach); 
\item a reweighting procedure 
      is performed in order to prevent the
      complete parton absorption in the medium ({\it reweighted} approach).
\end{Enumerate}

Our calculations contain only one free parameter that was adjusted to
the measured nuclear modification factor in central collisions at
$\sqrtsNN=200~\gev$ at RHIC.
The same  parameter was then employed to extract the centrality
dependence of the nuclear modification factor for hadrons and di-hadrons,
and the azimuthal anisotropy parameter $v_2$.
Some of these observables were simulated for $\sqrtsNN=62.4~\gev$ 
and $5.5~\tev$ collisions, scaling only the expected medium densities.

When comparing to experimental results, we observe a tendency for the outright
application of the theory, i.e.~{\it non-reweighted} approach, to fit
the $\RAA$ and $I_{\rm AA}$ centrality dependence reasonably well 
and, in general, better than the {\it reweighted} case.
The calculated value of $v_2$ lies about 2 standard deviations below 
the experimental data at $\pt\simeq 4$--$6~\gev$, suggesting the presence 
of collective elliptic flow effects up to these $\pt$ values; upcoming 
measurements with higher statistics
should be able to give a definitive answer whether 
elliptic flow still subsists at momenta larger than $7$--$8~\gev$ or not. 
We have also shown some 
predictions for the azimuthal variation of leading-particle suppression 
and jet-like correlations. 
We note that as a consequence of the 
azimuthal anisotropy for high-$\pt$ particles simulated in the present 
approach the low-energy particles `radiated' by quenched partons
might contribute a $v_2$ of the `opposite sign' in the low-$\pt$ region.
Thus, the parton quenching at the LHC could produce 
an apparent decrease of the elliptic flow at low $\pt$ with respect to 
the predictions of hydrodynamic calculations.

We observe that the {\it reweighting} `simulates' a softer transport
coefficient, i.e. 
it {\it de facto} allows for some partons to be emitted from the
away side, while a simple geometrical toy model excludes the away-side
partons once the model has been tuned on the measured $\RAA$. 
Furthermore, in the {\it reweighted} approach, $\RAA(\pt)$ is larger towards 
low transverse momenta, see e.g. Fig.~\ref{fig:RAAkCentral}, because the 
survival probability for a parton with energy $E$, 
$p_0/\int_0^E\d\epsilon\,P(\epsilon)$, 
increases when $E$ decreases. This feature appears to be 
unphysical, or at least non-intuitive.
Clearly, the full treatment of the difficulties encountered here should be 
tackled theoretically in a more complete way.

The inspection of the production-point distribution for energetic partons 
escaping the medium and of the average relative energy loss suffered 
by quarks and gluons in central collisions at top RHIC energy 
depicts the dense medium in the
nuclear overlap region as a black disk: either
the partons are absorbed or they escape from a thin shell close to the 
surface.

The present model, applied to the LHC, gives the interesting result that
the $\RAA$ value is essentially constant with $\pt$, and very low, 
up to the highest parton energies. As shown in Fig.~\ref{fig:RAAlhc},
this prediction differs substantially from others obtained for the 
LHC~\cite{vitev}.
Namely, in our model the black disk effect, which requires a large transport
coefficient, 
extends the strong suppression up to very high transverse momenta. 
This scenario would amount to decrease the number of high-energy jets 
by almost an order of magnitude and it should be considered in the 
future planning of experimental studies.